\documentclass[longauth]{aa}
\usepackage{graphicx}
\usepackage{placeins}
\usepackage[varg]{txfonts}
\usepackage{multirow}
\usepackage{float}
\usepackage{array}
\usepackage{lscape}
\bibpunct{(}{)}{;}{a}{}{,}
\usepackage{natbib,twoopt}
\usepackage[breaklinks=true]{hyperref}
\usepackage{amssymb,latexsym,graphicx,natbib,times,amsmath,xspace,ifthen, xcolor}
\usepackage[normalem]{ulem}
\usepackage{orcidlink}
\definecolor{linkblue}{rgb}{0,0.4,0.6}
\hypersetup{colorlinks, linkcolor={linkblue}, citecolor={linkblue}, urlcolor={linkblue}}

\defcitealias{BetAccGui23KS}{B23}

\title{The ALPINE-CRISTAL-JWST survey: spatially resolved star formation relations at $z\sim5$}

\author{
    C. Accard\inst{\ref{ObAS}}\orcidlink{0009-0005-9982-7239} 
    \and M. Béthermin\inst{\ref{ObAS}}\orcidlink{0000-0002-3915-2015}
    \and M. Boquien\inst{\ref{IAIChile}}\orcidlink{0000-0003-0946-6176}
    \and V. Buat\inst{\ref{LAM}}\orcidlink{0000-0003-3441-903X}
    \and L. Vallini\inst{\ref{INAFBologna}}\orcidlink{0000-0002-3258-3672}
    \and F. Renaud\inst{\ref{ObAS}}\orcidlink{0000-0001-5073-2267}
    \and K. Kraljic\inst{\ref{ObAS}}\orcidlink{0000-0001-6180-0245}
    \and M. Aravena\inst{\ref{IEAChile},\ref{MINGAL}}\orcidlink{0000-0002-6290-3198}
    \and P. Cassata\inst{\ref{DFAPadova},\ref{INAFPadova}}\orcidlink{0000-0002-6716-4400}
    \and E. da Cunha\inst{\ref{ICRAR}, \ref{ARC}}\orcidlink{0000-0001-9759-4797}
    \and P. Dam\inst{\ref{DFAPadova}}\orcidlink{0009-0007-7842-9930}
    \and I. de Looze\inst{\ref{Ghent}}\orcidlink{0000-0001-9419-6355}
    \and M. Dessauges-Zavadsky\inst{\ref{DAGeneva}}\orcidlink{0000-0003-0348-2917}
    \and Y. Dubois \inst{\ref{IAP}}\orcidlink{0000-0003-0225-6387}
    \and A. Faisst\inst{\ref{CalTech}} \orcidlink{0000-0002-9382-9832}
    \and Y. Fudamoto\inst{\ref{CFSJapan}}\orcidlink{0000-0001-7440-8832}
    \and M. Ginolfi\inst{\ref{DFAFirenze}, \ref{INAFFirenze}}\orcidlink{0000-0002-9122-1700}
    \and C. Gruppioni\inst{\ref{INAFBologna}}\orcidlink{0000-0002-5836-4056}
    \and S. Han\inst{\ref{IAP}}\orcidlink{0000-0001-9939-713X}
    \and R. Herrera-Camus\inst{\ref{MINGAL},\ref{DAUC}}\orcidlink{0000-0002-2775-0595}
    \and H. Inami\inst{\ref{Hiroshima}}\orcidlink{0000-0003-4268-0393}
    \and A.M. Koekemoer\inst{\ref{STSI}}
    \and B.C. Lemaux\inst{\ref{GeminiCal},\ref{GeminiCal2}}\orcidlink{0000-0002-1428-7036}
    \and J. Li\inst{\ref{ICRAR}, \ref{ARC}}\orcidlink{0000-0002-8184-5229}
    \and Y. Li\inst{\ref{Texas}}
    \and B. Mobasher\inst{\ref{DPAUC}}\orcidlink{0000-0001-5846-4404}
    \and J. Molina \inst{\ref{IFAValpChile},\ref{MINGAL}}\orcidlink{0000-0002-8136-8127}
    \and A. Nanni \inst{\ref{INAFTeramo},\ref{NCNRWarsaw}}\orcidlink{0000-0001-6652-1069}
    \and M. Palla\inst{\ref{INAFBologna},\ref{DFABologna}}\orcidlink{0000-0002-3574-9578}
    \and F. Pozzi\inst{\ref{INAFBologna}}\orcidlink{0000-0002-7412-647X}
    \and M. Relaño\inst{\ref{DFTCGrenada}}\orcidlink{0000-0003-1682-1148}
    \and M. Romano\inst{\ref{MPIBonn},\ref{INAFPadova}}\orcidlink{}
    \and P. Sawant\inst{\ref{NCNRWarsaw}}\orcidlink{0000-0002-0498-8074}
    \and J. Spilker\inst{\ref{Texas}}\orcidlink{0000-0003-3256-5615}
    \and A. Tsujita
    \and E. Veraldi\inst{\ref{SISSA}}\orcidlink{0009-0007-1304-7771}
    \and V. Villanueva\inst{\ref{UdeC}}\orcidlink{0000-0002-5877-379X}
    \and W. Wang\inst{\ref{CalTech}}
    \and S.K. Yi \inst{\ref{Yonsei}}\orcidlink{0000-0002-4556-2619}
    \and G. Zamorani\inst{\ref{INAFBologna}}\orcidlink{0000-0002-2318-301X}
}

\institute{
    Université de Strasbourg, CNRS, Observatoire Astronomique de Strasbourg, UMR 7550, 67000 Strasbourg, France \\ email: cedric.accard@astro.unistra.fr\label{ObAS}
    \and Université Côte d'Azur, Observatoire de la Côte d'Azur, CNRS, Laboratoire Lagrange, 06000, Nice, France \label{IAIChile} 
    \and Aix Marseille Univ, CNRS, CNES, LAM, Marseille, France\label{LAM} 
    \and INAF – Osservatorio di Astrofisica e Scienza dello Spazio di Bologna, Via Gobetti 93/3, 40129 Bologna, Italy\label{INAFBologna}
    \and Instituto de Estudios Astrof\'{\i}sicos, Facultad de Ingenier\'{\i}a y Ciencias, Universidad Diego Portales, Av. Ej\'ercito 441, Santiago, Chile\label{IEAChile}
    \and Millenium Nucleus for Galaxies (MINGAL), Concepción, Chile\label{MINGAL}
    \and Departamento de Astronomía, Universidad de Concepción, Barrio, Universitario, Concepción, Chile\label{DAUC}
    \and Dipartimento di Fisica e Astronomia, Università di Padova, Vicolo dell’Osservatorio 3, Padova, Italy \label{DFAPadova}
    \and INAF – Osservatorio Astronomico di Padova, Vicolo dell’ Osservatorio 5, 35122 Padova, Italy \label{INAFPadova}
    \and International Centre for Radio Astronomy Research (ICRAR), The University of Western Australia, M468, 35 Stirling Highway, Crawley, WA 6009, Australia \label{ICRAR}
    \and ARC Center of Excellence for All Sky Astrophysics in 3 Dimensions (ASTRO 3D), Australia \label{ARC}
    \and Sterrenkundig Observatorium, Ghent University, Krijgslaan 281 - S9, B9000 Ghent, Belgium\label{Ghent} 
    \and Department of Astronomy, University of Geneva, Chemin Pegasi 51, 1290 Versoix, Switzerland \label{DAGeneva}
    \and Institut d’Astrophysique de Paris, Sorbonne Université, CNRS, UMR 7095, 98 bis bd Arago, 75014 Paris, France\label{IAP}
    \and IPAC, California Institute of Technology, 1200 E California Boulevard, Pasadena, CA 91125, USA \label{CalTech} 
    \and Center for Frontier Science, Chiba University, 1-33 Yayoi-cho, Inage-ku, Chiba 263-8522, Japan \label{CFSJapan}
    \and Dipartimento di Fisica e Astronomia, Universitá degli Studi di Firenze, Via G. Sansone 1, 50019 Sesto Fiorentino, Firenze, Italy\label{DFAFirenze}
    \and INAF – Osservatorio Astrofisico di Arcetri, Largo E. Fermi 5, 50125 Firenze, Italy \label{INAFFirenze}
    \and Hiroshima Astrophysical Science Center, Hiroshima University, 1-3-1 Kagamiyama, Higashi-Hiroshima, Hiroshima 739-8526, Japan \label{Hiroshima}
    \and Space Telescope Science Institute, 3700 San Martin Drive, Baltimore, MD 21218, USA \label{STSI}
    \and Gemini Observatory, NSF NOIRLab, 670 N. A'ohoku Place, Hilo, Hawai'i, 96720, USA \label{GeminiCal}
    \and Department of Physics and Astronomy, University of California, Davis, One Shields Ave., Davis, CA 95616, USA \label{GeminiCal2}
    \and Department of Physics and Astronomy and George P. and Cynthia Woods Mitchell Institute for Fundamental Physics and Astronomy, Texas A\&M University, 4242 TAMU, College Station, TX 77843-4242, USA \label{Texas}
    \and Department of Physics and Astronomy, University of California, Riverside, 900 University Avenue, Riverside, CA 92521, USA \label{DPAUC}
    \and Instituto de F\'isica y Astronom\'ia, Universidad de Valpara\'iso, Avda. Gran Breta\~na 1111, Valpara\'iso, Chile \label{IFAValpChile}
    \and INAF - Osservatorio astronomico d’Abruzzo, Via Maggini SNC, 64100, Teramo, Italy \label{INAFTeramo}
    \and National Centre for Nuclear Research, ul. Pasteura 7, 02-093 Warsaw, Poland \label{NCNRWarsaw}
    \and Dipartimento di Fisica e Astronomia “Augusto Righi”, Alma Mater Studiorum, Università di Bologna, Via Gobetti 93/2, 40129 Bologna, Italy \label{DFABologna} 
    \and Dept. Fisica Teorica y del Cosmos, Universidad de Granada, Spain \label{DFTCGrenada}
    \and Max-Planck-Institut für Radioastronomie, Auf dem Hügel 69, 53121 Bonn, Germany \label{MPIBonn}
    \and Scuola Internazionale Superiore Studi Avanzati (SISSA), Physics Area, Via Bonomea 265, 34136 Trieste, Italy \label{SISSA}
    \and Departamento de Astronomía, Universidad de Concepción, Barrio Universitario, Concepción, Chile\label{UdeC}
    \and Department of Astronomy and Yonsei University Observatory, Yonsei University, Seoul 03722, Republic of Korea\label{Yonsei}
}
    
\date{Received 27/06/2025; accepted 15/08/2025}

\begin{document}
\abstract
{Star formation governs galaxy evolution, shaping stellar mass assembly and gas consumption across cosmic time. The Kennicutt–Schmidt (KS) relation, linking star formation rate (SFR) and gas surface densities, is fundamental to understand star formation regulation, yet remains poorly constrained at $z > 2$ due to observational limitations and uncertainties in locally calibrated gas tracers. The [CII] $158\,{\rm \mu m}$ line has recently emerged as a key probe of the cold ISM and star formation in the early Universe.}
{We investigate whether the resolved [CII]–SFR and KS relations established at low redshift remain valid at $4 < z < 6$ by analysing 13 main-sequence galaxies from the ALPINE and CRISTAL surveys, using multi-wavelength data (HST, JWST, ALMA) at $\sim2$\,kpc resolution.}
{We perform pixel-by-pixel spectral energy distribution (SED) modelling with CIGALE on resolution-homogenised images. We develop a statistical framework to fit the [CII]–SFR relation that accounts for pixel covariance and compare our results to classical fitting methods. We test two [CII]-to-gas conversion prescriptions to assess their impact on inferred gas surface densities and depletion times.}
{We find a resolved [CII]–SFR relation with a slope of $0.87 \pm 0.15$ and intrinsic scatter of $0.19 \pm 0.03$\,dex, which is shallower and tighter than previous studies at $z\sim5$. The resolved KS relation is highly sensitive to the [CII]-to-gas conversion factor: using a fixed global $\alpha_{\rm [CII]}$ yields depletion times of $0.5$–$1$\,Gyr, while a surface brightness-dependent $W_{\rm [CII]}$, accounting for local ISM conditions, places some galaxies with high gas density in the starburst regime ($<0.1$\,Gyr).}
{Future inputs from both simulations and observations are required to better understand how the [CII]-to-gas conversion factor depends on local ISM properties. We need to break this fundamental limit to properly study the KS relation at $z\gtrsim4$.}


\maketitle

\section{Introduction}\label{sec:intro}

The study of star formation in the early Universe ($z>2$) is a cornerstone of modern astrophysics, crucial for understanding galaxy evolution, mass assembly, and cosmic star formation history \citep[e.g.][]{Madau2014, Freundlich2024}. The Kennicutt-Schmidt relation, linking star formation rate surface density ($\Sigma_\mathrm{SFR}$) to gas surface density ($\Sigma_\mathrm{gas}$) through a power law, has been fundamental for understanding the conversion of the gas into stars across cosmic time \citep[e.g.][]{kennicutt1998global, delosReyes_2019}. 
Initially proposed by \citet{1959Schmidt} in terms of volume densities within the Milky Way, this relation was later reformulated for extragalactic studies using measurements of surface densities \citep{kennicutt1998global}. While extensively studied locally \citep[e.g.][]{Bigiel2008, KennicuttEvans2012, Leroy2013, Pessa2021, Jimenez2023}, its high-redshift behaviour ($z\gtrsim2$) remains poorly constrained with significant uncertainties in both the slope and intrinsic scatter of the relation. These uncertainties arise primarily from low spatial resolution, limited sensitivity \citep[e.g.][]{Freundlich2013}, and sample representativeness issues \citep[e.g.][]{Vallini2024}, hindering progress towards an understanding of early-Universe star formation. Thanks to existing observational facilities such as Hubble Space Telescope (HST) and Atacama large millimetre/submillimetre array (ALMA), and the recent advancements made possible with the James Webb Space Telescope (JWST), we can now perform multi-wavelength analysis to probe galaxies at unprecedented redshifts and resolutions \citep[e.g.][]{Nagy, Hashimoto_2023, Wang_2024, Zavala_2024}. Those observations have opened new avenues for investigating the physical processes governing star formation in the first few billion years after the Big Bang.

Traditional molecular gas tracers, such as CO, become increasingly difficult to detect at higher redshift \citep[$z>2$, e.g.][]{Molina2019, Tacconi2020}, creating a critical gap in our understanding of early-Universe star formation \citep[e.g.][]{Genzel10SFrelations, Scoville2016}. This limitation has been partially overcome by the use of the [CII] $158\,{\rm \mu m}$ emission line, now regularly used as an alternative tracer of molecular gas \citep{ZanellaCII}. That fine-structure line is typically one of the most luminous far-infrared line in star-forming galaxies and is redshifted into ALMA's atmospheric windows at $z\gtrsim4$. This provides unique insights into cold gas kinematics and is found to correlate with star formation \citep[e.g.][]{DeLooze2014, Herrera-Camus_2015}. In particular, \citet{Herrera-Camus_2015} discuss how [CII], as the main coolant of photo-dissociation regions (PDRs; see \citealt{Wolfire2022}), is physically linked to the SFR through the photoelectric effect, which couples the heating of the gas to the presence of young, massive stars. Several theoretical and observational studies have demonstrated that [CII] emission could also be strongly linked to molecular gas mass, particularly in the local Universe \citep[e.g.][]{ZanellaCII, Madden20CII, DEugenio23}. However, recent work at high redshift suggests a more complex scenario. For instance, \citet{Dessauges2020} find that, in main-sequence galaxies at $z\sim5$, the gas masses inferred from [CII] luminosity are comparable to those derived from dynamical mass and far-infrared continuum estimates \citep{Scoville2014}. This implies that either the total gas reservoir in these galaxies is predominantly molecular, or that [CII] traces the total gas mass, including both molecular and atomic components, rather than exclusively the molecular phase. The complex origin of [CII] emission, arising from both PDRs and diffuse ionised gas, further complicates its interpretation, necessitating careful analysis of observational data \citep[e.g.][]{Vallini_2015, Wolfire2022}

The ALMA large program to investigate C+ at early times (ALPINE; \citealt{Lefevre20ALPINE, bethermin2020alpine, FaisstReviewALPINE}) has significantly expanded our sample of [CII]-detected galaxies at $4 < z < 6$, offering a first opportunity to study star formation relations in main-sequence galaxies during this epoch. This survey, comprising 118 star-forming galaxies, has already yielded important insights into the interstellar medium (ISM) properties \citep{Vanderhoof2022, Veraldi2025} and star formation characteristics of galaxies in the early Universe \citep{Schaerer2020, FaisstReviewALPINE, Romano2022}. Follow-up studies combining higher resolution of four ALPINE galaxies with HST observations have allowed for a more detailed view of individual galaxy properties. The new ALMA observations, allowing to resolve the spatial extent of [CII] emission on kpc scales, provide information about their morpho-kinematics \citep{TobiMorpho} and the distribution of star formation and cold gas linked through the KS relation \citep[B23 in what follows]{BetAccGui23KS}. The latter shows that, despite their disturbed morphologies, likely due to mergers, the studied galaxies appeared to follow an extension of the local KS relation at higher gas surface densities. These studies have so far been limited to small sample sizes, which limits the statistical significance of their conclusions. To overcome this limitation, the [CII] resolved ISM in star-forming galaxies with ALMA (CRISTAL) Large Program (2021.1.00280.L, \citealt{herreracamus2025almacristalsurveygasdust}) was initiated as an ALMA Cycle 8 follow-up to the ALPINE survey. The program aims at obtaining spatially-resolved images of 19 [CII]-detected galaxies at $4.430\leq z \leq 5.689$ with existing HST rest-frame UV emission images. These galaxies were selected based on their specific star formation rate (sSFR) within a factor of three of the star-forming main sequence at their respective redshifts, and with stellar masses $M_\star \geq 10^{9.5}\,M_\odot$.

Leveraging the expanded capabilities of the ALPINE and CRISTAL surveys, we present an analysis of 13 main-sequence galaxies at z $\sim$ 5, combining high-resolution data from JWST, HST, and ALMA. This multi-wavelength approach probes both dust-obscured and unobscured star formation, as well as the cold gas component traced by [CII]. The inclusion of JWST data significantly enhances our spectral coverage, enabling panchromatic rest frame UV-to-radio Spectral energy distribution (SED) modelling for each line of sight, which was previously unattainable with HST and ALMA data alone. This is performed using the python-based code investigating galaxy emission (CIGALE; \citealt{boquien2019cigale}). With the production of spatially resolved maps below the kpc scale of inferred physical properties, we can investigate the slope and scatter of $\Sigma_\mathrm{[CII]}-\Sigma_\mathrm{SFR}$ relation in the spatially resolved manner. We adopt a comprehensive statistical approach to handle uncertainties of the different measurements and estimates, which will allow us to derive the slope and intrinsic scatter of the [CII]-SFR relation. We explore the impact of different [CII]-to-gas conversion factors on estimates of gas masses and star formation efficiencies via the KS relation. 

The goal of this work is to determine whether the resolved [CII]–SFR and KS relations established at low redshift remain valid at $z\gtrsim4$, or if evolving ISM conditions and gas density regimes at high redshift significantly impact the link between cold gas and star formation. The structure of this paper is as follows. In Sect.\,\ref{sec:Data}, we describe the observations, sample selection, and data homogenisation procedures. Sect.\,\ref{sec:methods} details our methodology, including the resolved SED fitting and the statistical treatment of uncertainties. Our main results, including the resolved [CII]-SFR relation and the impact of [CII]-to-gas conversion factors on the KS relation, are presented in Sect.\,\ref{sec:results}. We discuss the implications of our findings in Sect.\,\ref{sec:discussion}, and summarise our conclusions in Sect.\,\ref{sec:conclusion}.

To ensure consistency with past works on the ALPINE survey, we assume a flat $\mathrm{\Lambda}$CDM cosmology (h = 0.7, $\Omega_\mathrm{\Lambda}$ = 0.7, $\Omega_\mathrm{m}$ = 0.3) and a Chabrier initial mass function (IMF, \citealt{Chabrier}).

\section{Observations and data homogenisation}\label{sec:Data}
\subsection{Sample selection}\label{sec:sampleselection}

\begin{table*}
\caption{Summary of the ALPINE-CRISTAL sample used in this work.}
\label{table:sources}
\centering
\begin{tabular}{l l l l l l l l}
\hline\hline
\vspace{-0.8em}\\
Original name & CRISTAL ID & $z_\mathrm{[CII]}$ & $\alpha_\mathrm{J2000}$ & $\delta_\mathrm{J2000}$ & log$_{10}$(M$_\star$) & log$_{10}$(SFR)& Beam size \\
& & & deg & deg & M$_\sun$ & M$_\sun$\,yr$^{-1}$ & $\rm arcsec \times arcsec$\\\vspace{-1em}\\\vspace{-1em}\\\hline
\vspace{-0.8em}\\
\multicolumn{8}{c}{CRISTAL Large Program} \\\hline
\vspace{-0.8em}\\
    DEIMOS\_COSMOS\_842313    & CRISTAL-01      & 4.530 & 150.2271 & +2.5762 & $10.65\pm0.50$ & $2.31\pm0.74$ & $0.18\times0.18$\\
    DEIMOS\_COSMOS\_848185    & CRISTAL-02      & 5.294 & 150.0896 & +2.5864 & $10.30\pm0.23$ & $1.71\pm0.29$ & $0.33\times0.28$\\
    DEIMOS\_COSMOS\_536534    & CRISTAL-03      & 5.689 & 149.9719 & +2.1182 & $10.40\pm0.29$ & $1.79\pm0.31$ & $0.51\times0.40$\\
    vuds\_cosmos\_5100822662  & CRISTAL-04a     & 4.520 & 149.7413 & +2.0809 & $10.15\pm0.29$ & $1.89\pm0.21$ & $0.42\times0.35$\\
    vuds\_cosmos\_5100541407  & CRISTAL-06a     & 4.562 & 150.2538 & +1.8094 & $10.09\pm0.30$ & $1.62\pm0.34$ & $0.37\times0.28$\\
    DEIMOS\_COSMOS\_873321    & CRISTAL-07ab    & 5.154 & 150.0166 & +2.6266 & $10.00\pm0.33$ & $1.89\pm0.26$ & $0.45\times0.28$\\
    DEIMOS\_COSMOS\_519281    & CRISTAL-09a     & 5.575 & 149.7537 & +2.0910 & $9.85\pm0.39$ & $1.51\pm0.32$ & $0.18\times0.16$\\
    DEIMOS\_COSMOS\_630594    & CRISTAL-11      & 4.439 & 150.1358 & +2.2579 & $9.68\pm0.33$ & $1.57\pm0.31$ & $0.25\times0.21$\\
    vuds\_cosmos\_5100994794  & CRISTAL-13      & 4.579 & 150.1715 & +2.2873 & $9.65\pm0.34$ & $1.51\pm0.41$ & $0.18\times0.14$\\
    vuds\_cosmos\_5101244930  & CRISTAL-15      & 4.580 & 150.1986 & +2.3006 & $9.69\pm0.33$ & $1.44\pm0.24$ & $0.23\times0.21$\\
    DEIMOS\_COSMOS\_494763    & CRISTAL-19      & 5.233 & 150.0213 & +2.0534 & $9.51\pm0.36$ & $1.45\pm0.36$ & $0.39\times0.29$\\
\vspace{-0.8em}\\
\hline
\vspace{-0.8em}\\
\multicolumn{8}{c}{ALPINE follow-up} \\
\vspace{-1em}\\
\hline
\vspace{-0.8em}\\
    DEIMOS\_COSMOS\_873756    & CRISTAL-24      & 4.546 & 150.0113 & +2.6278 & $10.53\pm0.08$ & $2.06\pm0.22$ & $0.22\times0.15$\\
    vuds\_cosmos\_5110377875  & None\tablefootmark{a} & 4.551 & 150.3847 & +2.4084 & $10.16\pm0.21$ & $1.99\pm0.26$ & $0.40\times0.35$\\\vspace{-1em}\\
\hline
\end{tabular}
\tablefoot{The first column presents the original COSMOS survey names, while the second column shows CRISTAL IDs following the naming convention in \citet{Mitsuhashi_2024}. The \(\mathrm{z}_\mathrm{[CII]}\) column lists redshifts derived from ALMA [C\,\textsc{ii}] line observations of the CRISTAL survey \citep{herreracamus2025almacristalsurveygasdust}, for which formal uncertainty estimates are not provided. Nonetheless, these values are consistent with those from \citet{bethermin2020alpine}, who report typical uncertainties of approximately 0.0005. Coordinates are given in J2000 epoch. Integrated mass and SFR are those reported in \cite{herreracamus2025almacristalsurveygasdust} apart from VC875 whose values are from the \cite{bethermin2020alpine} catalogue. Beam sizes of the ALMA observations are reported as major axis $\times$ minor axis in arcseconds. \tablefoottext{a}{VC875 has not been used in previous CRISTAL work, therefore it has no CRISTAL ID.}}
\end{table*}

For this study, we select a subset of 11 galaxies from the CRISTAL survey with homogeneous ALMA, HST, and JWST observations, that is same filter coverage across all sources: HST/ACS F814W, JWST/NIRCam F115W, F150W, F277W, and F444W, along with ALMA [CII] line and associated dust continuum observations. As ALMA provides the coarsest spatial resolution among these datasets, the effective resolution of our analysis is set by the ALMA beam (see Sect.\,\ref{sec:Observations}). This consistent multi-wavelength coverage enables the application of a uniform data homogenisation pipeline across the entire sample. In addition to the CRISTAL objects, we considered another galaxy from the ALPINE survey re-observed as part of the ALMA project 2019.1.00226.S (PI: E. Ibar). Although not part of the original CRISTAL survey, this galaxy (DEIMOS\_COSMOS\_873756) was assigned a CRISTAL identifier (CRISTAL\_24). Finally, we included vuds\_cosmos\_5110377875 (hereafter VC875) from the ALPINE survey, as it meets our selection criteria (ALMA project 2022.1.01118.S, PI: M. Béthermin).

The final sample contains 13 star-forming main-sequence galaxies between $z=4.439$ and $z=5.689$, all located in the Cosmic Evolution Survey field (COSMOS; \citealt{Scoville2007}). Accurate redshifts for these galaxies have been computed from extensive optical spectroscopy \citep{LeFevre2015,Hasinger2018} and then confirmed and refined with [CII] ALPINE data \citep{bethermin2020alpine}. A summary of the final sample is found in Table \ref{table:sources}.

\subsection{Observations}\label{sec:Observations}

ALMA observations were conducted in Band 7, targeting the $158\,{\rm \mu m}$ rest-frame [CII] line ($\nu_{rest} = 1900.54\,{\rm GHz}$)  using various antenna configurations. For the CRISTAL survey objects, the final data combined extended (C43-5 or C43-6; typical beam sizes $0.1\arcsec$--$0.3\arcsec$ and spatial scales of $1.2$--$2.0$\,kpc) and compact array (C43-1 or C43-2; $0.7\arcsec$--$1.0\arcsec$, $4.5$--$6$\,kpc) configurations, incorporating available ancillary data. CRISTAL\_24 was observed using C43-3 ($\sim 0.4 \arcsec$, $\sim 2.7$\,kpc, medium resolution) and C43-5 ($\sim 0.2 \arcsec$, $\sim 1.5$\,kpc, high resolution) configurations. The CRISTAL team consistently reprocessed these data to produce [CII] moment and dust continuum emission maps using Briggs weighting (see Sect.\,5 of \citealt{herreracamus2025almacristalsurveygasdust}). VC875 was observed in the C43-4 configuration ($\sim 0.3 \arcsec$, $\sim2.0$\,kpc) and reduced independently \citep{BetAccGui23KS}.

HST observations are cutouts from the COSMOS survey \citep{Koekemoer2007}, obtained with the HST ACS camera and the F814W filter, at a FWHM resolution of $0.13 \arcsec$. At the considered redshifts, this corresponds to rest-frame emission of $120$--$147$\,nm and spatial scales of $0.76$--$0.86$\,kpc.

All sources were observed with four JWST/NIRCam filters: F115W, F150W, F277W, and F444W, probing rest-frame emissions at $183$--$211$\,nm, $238$--$276$\,nm, $440$--$509$\,nm, and $705$--$816$\,nm, respectively. The achieved FWHM resolutions are $0.037 \arcsec,~0.049 \arcsec,~0.088 \arcsec, \text{ and } 0.140 \arcsec$; respectively giving spatial scales of $0.21$--$0.24$\,kpc, $0.29$--$0.33$\,kpc, $0.52$--$0.59$\,kpc, $0.82$--$0.93$\,kpc, for our redshift range. These observations were part of the COSMOS-Web program (PID: 1727; co-PIs: Casey \& Kartaltepe; \citealt{COSMOS_WEB}).

\begin{figure*}
\centering
     \includegraphics[width=\textwidth]{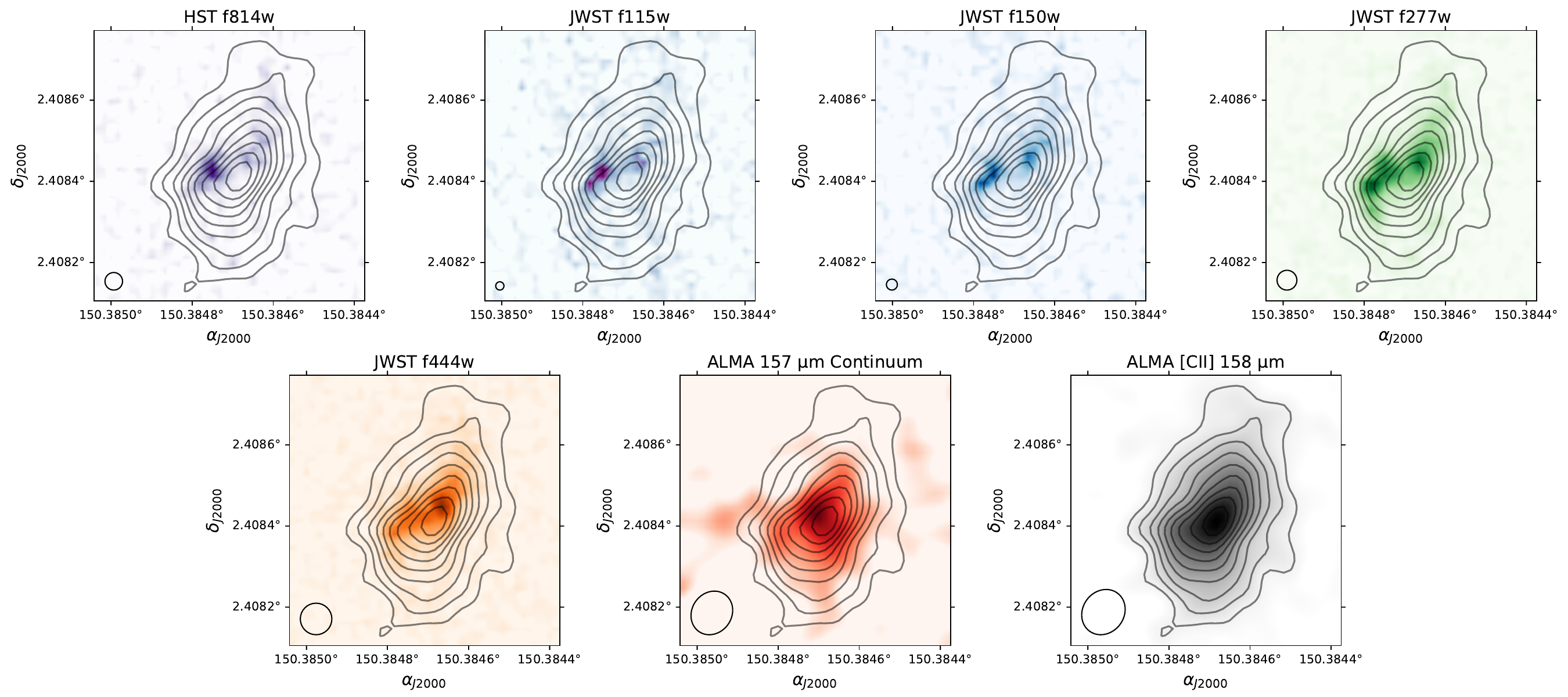}
     \caption{Multi-wavelength data for the galaxy VC875, shown at their native resolutions. From left to right : The first row presents HST/ACS F814W (rest-frame 145 nm), JWST/NIRCam F115W (207 nm), F150W (270 nm), and F277W (499 nm). The second row shows JWST/F444W (800 nm), ALMA dust continuum emission around 157 µm, and ALMA [CII] $158\,{\rm \mu m}$ emission. The contours are those of the ALMA [CII] line emission maps starting from $3\sigma$ increasing by steps of $2\sigma$. Ellipses in the bottom-left corners represent the central lobe of the PSF for HST and JWST images, and the synthesised beam for ALMA images.}
     \label{fig:VC875_native}
\end{figure*}

An example of the complete multi-wavelength coverage available for our sample is presented in Fig.\,\ref{fig:VC875_native}, for VC875. VC875 shows two UV-bright components in HST F814W and $\lambda < 3\,\mu$m JWST bands, but the JWST/F444W and ALMA bands, as well as the stellar mass map from CIGALE (see Fig.\,\ref{fig:CIGALEparams}), reveal a single component, supporting its treatment as a single galaxy with UV-bright clumps.

\subsection{Homogenisation of the resolution}\label{sec:ResHom}

To perform accurate pixel-by-pixel comparisons across all available bands for our SED modelling process, we need to ensure that the information contained within each pixel corresponds to the same region of the sky, for all observational bands. This consistency is crucial for deriving reliable physical properties from multi-wavelength data. Therefore, we perform point spread function (PSF) homogenisation across all bands to match their resolutions. We use the ALMA beam as the target resolution since it is the coarsest in our sample. HST and JWST PSFs central lobes are modelled with 2D Gaussian functions to retrieve their full width at half maximum and maximum intensity. 

Cleaned synthesised ALMA beams models are created based on dust continuum map header parameters, assuming a two-dimensional elliptical Gaussian shape. Convolution kernels are generated to transform HST and JWST PSFs to match the target ALMA beam taking into account different grid orientations, pixel scales, and normalisation conventions. 
We reproject HST and JWST convolved images onto the ALMA astrometric grid using the \texttt{reproject\_interp} module from the python package \texttt{reproject}\footnote{\url{https://reproject.readthedocs.io}}, ensuring consistent pixel sampling and astrometric alignment across all bands. Applying the procedure to a point-like source with unit flux in HST/JWST native resolution, we find the reconstructed average flux to be 2.6\% lower than the expected peak value of unity across the different filters. The missing flux likely resides in the secondary lobes not taken into account, and we consider this effect to be negligible compared to our overall flux calibration uncertainties.

\begin{figure}
\centering
     \includegraphics[width=88mm]{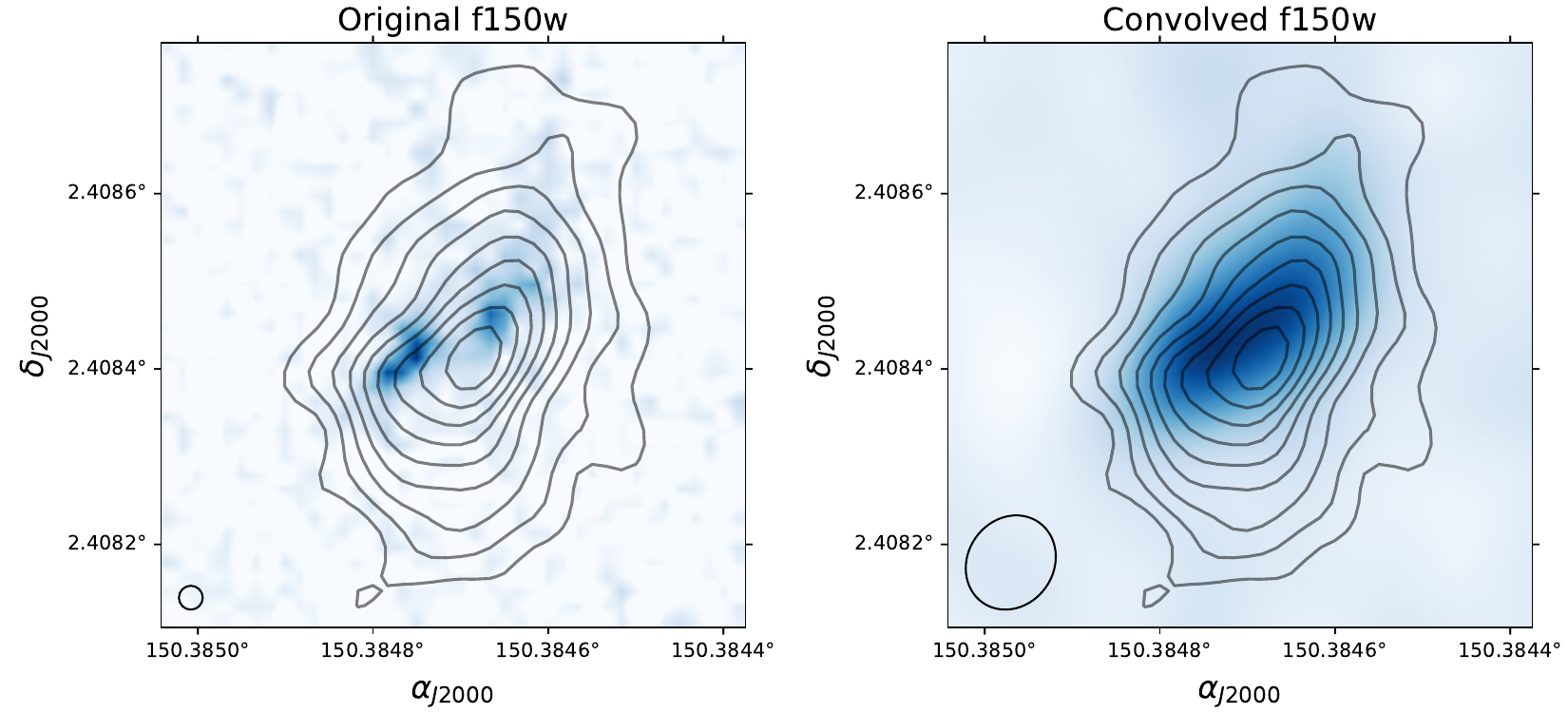}
     \caption{Native resolution image of VC875 (left panel) and the homogenised version matching the ALMA resolution (right panel) for the JWST/NIRCam F150W filter. The contours and ellipses are the same as in Fig.\,\ref{fig:VC875_native}.}
     \label{fig:VC875comp}
\end{figure}

Figure\,\ref{fig:VC875comp} shows the effect of the resolution homogenisation of the JWST/NIRCam F150W filter in the case of VC875. The complete set of homogenised observations for this source is found in \ref{fig:VC875smoothed}. As a result of this resolution homogenisation, we achieve an average beam size of $0.3\arcsec$ ($\sim$2\,kpc) across the sample. Although the pixel scale is $\sim$0.2\,kpc, the physical information is correlated on the scale of the beam. We take into account these correlations induced by our resolution homogenisation in our analysis of the [CII]-SFR relation (see Sect.\,\ref{sec:resampling}).

\subsection{Photometric and line flux unit conversion}\label{sec:UnitsHom}

To prepare the data for SED modelling, we convert all photometric data to $\mu\mathrm{Jy\,beam}^{-1}$, and ALMA dust continuum emission flux to dust luminosity. Here "beam" refers to the synthesised ALMA beam, not the PSF of each instrument. Expressing fluxes per beam ensures that the measurements are referenced to the area over which the instrument is sensitive. The dust luminosity $L_\mathrm{IR}$ ($\mathrm{W\,beam}^{-1}$) is calculated from the monochromatic flux $S_\nu$ (Jy $\rm beam^{-1}$) using:
\begin{equation}
    L_{\mathrm{IR}} = 10^{-26} \times \frac{1}{0.13}\,\nu_{\mathrm{em}}\,\frac{4\pi D_L^2}{1+z}\, S_{\nu},
\end{equation}
where $\nu_{\mathrm{em}}$ is the rest-frame emission frequency at 158 micron (Hz), $D_L$ is the luminosity distance (m), and $z$ is the redshift. The factor $10^{-26}$ converts Jy to $\mathrm{W\,m^{-2}\,Hz^{-1}}$, while the $1/(1+z)$ term accounts for cosmological redshift. 
We adopt the empirical conversion factor of \(0.13 \pm 0.02\) from \cite{Khusanova} to convert the \(158\,\mathrm{\mu m}\) rest-frame monochromatic luminosity into the total infrared luminosity \(L_{\mathrm{IR}}\). This value corresponds to the calibration for the redshift bin \({4 < z < 5}\) and is directly applicable to our sample, which predominantly lies within this range. The other sources are in the $5 < z < 6$ bin, in which \citet{Khusanova} measured a conversion factor of \(0.12 \pm 0.03\). Since the two values are compatible with no redshift dependence, we use the more accurate value found at $4 < z < 5$ for the full sample. Both factors were calibrated using stacked SEDs. A compilation of SED templates from the literature was fitted to these SEDs and only the good fits were kept (reduced $\chi^2<1.5$). The conversion factor was then estimated using the mean and dispersion of these templates. The typical dust temperatures corresponding to these stacked SEDs are \(T_{\mathrm{d}} = 41 \pm 1\,\mathrm{K}\) at \(z \sim 4.5\) and \(T_{\mathrm{d}} = 43 \pm 5\,\mathrm{K}\) at \(z \sim 5.5\) for galaxies with \(\mathrm{SFR} > 10\,M_\odot\,\mathrm{yr}^{-1}\) \citep[see Appendix A of][]{Khusanova}. We note that adopting a constant conversion factor assumes a uniform dust temperature across the sample, and variations could introduce systematic effects on \(L_{\mathrm{IR}}\) and thus \(\Sigma_{\mathrm{SFR}}\) estimates \citep{Faisst_2017,Cochrane_2022}. We include this IR dust luminosity in the SED modelling process as a constraint on dust attenuation models (see Sect.\,\ref{sec:SEDfit}).

The [CII] surface brightness ($\Sigma_{\mathrm{[CII]}}$ in $\mathrm{L}_\sun$ $\mathrm{kpc}^{-2}$) is derived from the [CII] moment-0 map ($m_\mathrm{[CII]}$ in Jy km $\mathrm{s}^{-1}$ $\mathrm{beam}^{-1}$) using:
\begin{equation}
    \Sigma_{\mathrm{[CII]}} =  \frac{m_\mathrm{[CII]}}{D_\mathrm{A}^2\, \Omega_{\mathrm{Beam}}} \, \left(1.04 \times 10^{-3}\,\frac{\mathrm{L}_\sun}{\mathrm{GHz\,Mpc}^2\, \mathrm{Jy\,km\,s}^{-1}} \right) \, D_\mathrm{L}^2 \, \nu_{\mathrm{obs}},
\end{equation}
where $D_A^2\,\Omega_{\mathrm{Beam}}$ is the physical area associated with the synthesised beam, defined by $D_\mathrm{A}$, the angular distance (in kpc), and $\Omega_{\mathrm{Beam}}$, the solid angle of the beam. The remaining factors correspond to the conversion from line flux to luminosity \citep{Solomon92}, where $D_\mathrm{L}$ is the luminosity distance (in Mpc) and $\nu_{\mathrm{obs}}$ the observed frequency (in GHz). 

\subsection{Uncertainty estimation}\label{sec:ObsMocks}

For HST and JWST images, we incorporate the convolution systematic error from Section \ref{sec:ResHom}. In addition, we apply a relative uncertainty of 3\% to JWST images to account for potential calibration uncertainties \citep{ErrBudJWST}. The conversion from  continuum flux to $L_\mathrm{IR}$ relies on an empirical factor of $0.13 \pm 0.02$, which introduces a relative systematic uncertainty of approximately 15\% ($0.02/0.13$) to the derived $L_\mathrm{IR}$ values in our continuum images. For images lacking error maps (HST data and $L_\mathrm{IR}$), we assume white noise at the native resolution and estimate the measurement uncertainty using the standard deviation of the noise in the convolved then reprojected image. For $\Sigma_\mathrm{[CII]}$, we use the pixel-by-pixel error maps provided by the CRISTAL team. 

JWST maps are provided with their associated error maps. We produce variance maps of the resolution-homogenised maps (in Jy/pixel units) by convolving the noise variance map (error map squared) by the kernel: \begin{equation}
\sigma_{\rm conv}^2 = \sigma_{\rm pixel}^2 \ast K^2,
\end{equation}
where $\sigma_{\rm conv}^2$ and $\sigma_{\rm pixel}^2$ are the noise variance of the convolved and original maps respectively, and K is the homogenisation kernel. We compare the results with the standard deviation of the noise in an empty area of the homogenised map, and we find a small excess likely due to a small contribution of correlated noise already present in the original maps and not taken into account by our computation. This excess remains negligible for the short-wavelength filters (less than 6\% for F115W and F150W), but becomes increasingly significant at longer wavelengths, reaching up to 14\% for F277W and F444W. We thus apply a scaling factor (1--1.14, depending on the source and band) to the resolution-homogenised error maps to correct for this effect.

To account for this error budget in our analysis, we generate 100 perturbed realisations of each image. This process involves adding white noise with a standard deviation equal to the measured pixel-level noise at the native resolution of each image, followed by the application of systematic errors. HST and JWST perturbed images are subsequently convolved. This method accounts for the non-independence of the pixels from spatially-correlated noise and the smoothing of the maps.

\section{Methods}\label{sec:methods}
\subsection{Resolved SED fitting}\label{sec:SEDfit}

\begin{figure*}
    \centering
    \includegraphics[width=0.966\textwidth]{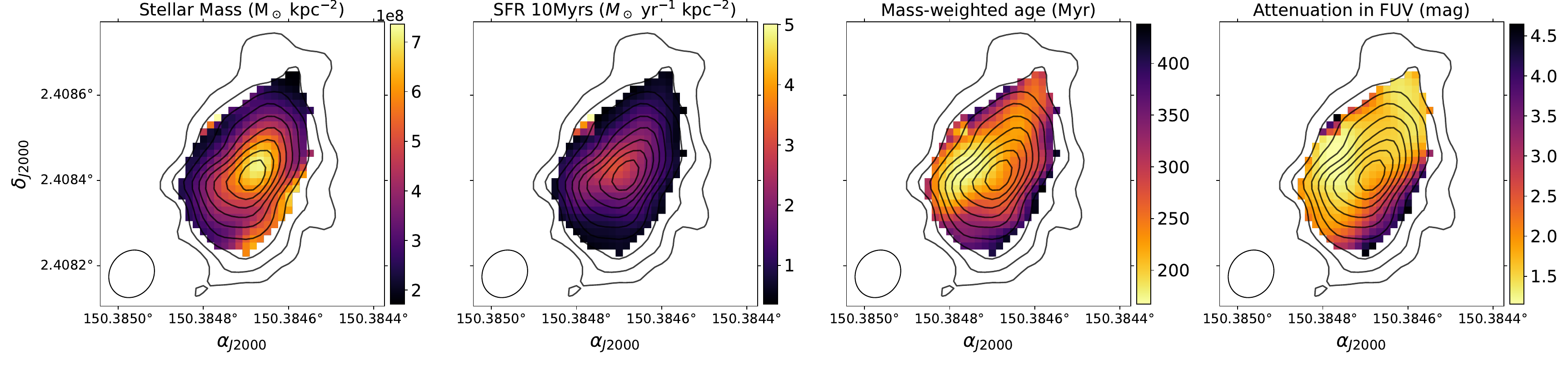}
    \caption{Example of physical parameters maps obtained with the CIGALE SED modelling for VC875. From left to right : stellar mass surface density ($\mathrm{M}_\odot\,\mathrm{kpc}^{-2}$), star formation rate surface density averaged over 10\,Myrs ($\mathrm{M}_\odot\,\mathrm{yr}^{-1}\,\mathrm{kpc}^{-2}$), mass-weighted age (Myr) and attenuation in FUV rest-frame (mag). The contours are those of the [CII] luminosity map, starting at $3\sigma$ and increasing by steps of $2\sigma$. The ellipse in the bottom left of each panel represent the synthesised ALMA beam. The maps are shown for the pixels having $L_{\rm [CII]}\geq5\sigma_\mathrm{[CII]}$ and S/N$\geq2$ in at least three bands.}
    \label{fig:CIGALEparams}    
\end{figure*}

Our primary objective is to derive the SFR along each line of sight for the galaxies in our sample by combining multi-wavelength observations. At a spatial resolution of 2\,kpc, the use of an energy-conserving SED modelling code is justified, as the energy balance assumption-where UV/optical energy absorbed by dust is re-emitted in the IR-remains physically meaningful at this scale \citep[e.g.][]{Boquien2015}.

To this end, we use CIGALE, a versatile Python tool capable of modelling galaxy emission from X-ray to radio wavelengths \citep{boquien2019cigale}. Its modular design allows for the inclusion of diverse star formation histories (SFH), dust attenuation laws, and emission line models tailored to specific scientific objectives. CIGALE uses a Bayesian approach to perform parameter estimation, generating a large library of theoretical SEDs based on user-defined models. Observed data are then compared to these models, and posterior probability distributions are calculated for each parameter. This Bayesian framework ensures that uncertainties and degeneracies between parameters are properly propagated into the resulting parameter estimates. 

In this section, we present the SED modelling parameters used for our analysis. We follow \citet{BoquienParam}, who analysed integrated measurements of the ALPINE galaxies (our parent sample), and \citet{BuatParam}, who focused on dust-rich galaxies at $z\sim2$. We refer the reader to those works for a comprehensive discussion of parameter choices. Our focus here is to detail the differences. A summary of our parameters is given in Table \ref{tab:Param}.

We adopt delayed SFH model with an additional recent variation, as described by \citet{CieslaBQ}. In this framework, star formation begins after an initial delay corresponding to the time between the Big Bang and the onset of the first stars (here set to 250 Myr). The main stellar population forms according to a delayed SFH, which rises linearly from the onset, peaks at a characteristic timescale, and then declines exponentially. To capture recent deviations from this smooth evolution, we introduce a modification to the SFR at a specified time in the recent past. The SFH is thus expressed as:
\begin{equation}
\mathrm{SFR}(t) =
\begin{cases}
t \, \exp\left(-\frac{t}{\tau}\right), & t < t_\mathrm{var} \\
r_{\rm SFR} \, t \, \exp\left(-\frac{t}{\tau}\right), & t \geq t_\mathrm{var}
\end{cases}
\end{equation}

where $t$ is the time since the onset of star formation, $\tau$ is the SFH timescale, $t_\mathrm{var}$ is the time since the onset of the recent SFR variation, and $r_{\rm SFR}$ is the factor by which the SFR is increased or decreased. This approach provides flexibility to model both the overall stellar mass assembly and any recent changes in star formation activity that a simple delayed SFH model would not account for.

For all sources, we adopt the two-component dust attenuation model of \citet{CF00}. This choice is motivated by its physical basis and agreement with radiative transfer predictions, allowing it to be applicable across the full range of dust attenuation, from weakly to highly obscured regions. While the modified \citet{Calz1994,Calz2000} relation \citep{Noll2009} is often used for star-forming galaxies, it is less appropriate for low attenuation or highly obscured systems \citep[e.g.][]{BoquienParam}, and allowing its slope parameter ($\delta$) to vary could lead to overfitting given our limited number of photometric bands. Although most of our sample shows no significant differences between the two models (as illustrated by VC875; see Appendix\,\ref{App:SFRvsSFR}), the \citet{CF00} law provides a substantially better agreement between SFR estimators for CRISTAL\_24 (see Fig.\,\ref{fig:C24SFRcomp}), further consolidating our choice of attenuation law.

We also evaluate the impact of including the ALMA-derived total infrared luminosity ($L_\mathrm{IR}$, computed from the $158\,{\rm \mu m}$ continuum in Sect.\,\ref{sec:UnitsHom}) as a dust luminosity constraint in our SED modelling. We find that omitting this constraint causes the models to systematically overestimate the SFR compared to estimates from combined UV and far-IR data. This finding is consistent with \citet{Juno}, who reported a similar discrepancy in an overlapping sample. Their analysis, using region-averaged measurements with the $\texttt{magphys}$ \citep{magphys} SED fitting code, likewise highlights the necessity of the dust luminosity constraint in SED modelling. Our results, derived from a pixel-based CIGALE approach, reinforce this conclusion.

In Appendix\,\ref{App:SFRvsSFR}, we discuss our final modelling choices in detail, using the galaxy VC875 as a case study. These choices include the dust luminosity constraint, the adopted dust attenuation law, and the treatment of recent star formation episodes. A similar analysis was performed for all other sources in the sample.

To ensure robust measurements, we use only pixels with $\Sigma_{\rm [CII]}\geq5\sigma_{\rm [CII]}$ and S/N$\geq2$ in at least three bands. The latter cut, applied to the data used for SFR estimates, excludes only about 6\% of pixels that passed the $L_{\rm [CII]}$ threshold, confirming that our selection is primarily driven by [CII] detectability. This ensures the robustness of the [CII] measurements while minimising biases in the inferred SFR distribution. 

SFR uncertainties are estimated from the 16th, 50th, and 84th percentiles of the log(SFR) distributions obtained via resampling noise realisations (see Sect.\,\ref{sec:ObsMocks}). This approach accounts for pixel correlations introduced by resolution-matching. It avoids biases from assuming Gaussian errors, as the SFR distributions are typically skewed.

Figure\,\ref{fig:CIGALEparams} presents the maps of stellar mass, SFR, mass-weighted age, and FUV attenuation for VC875, as derived from SED modelling after applying all pixel selection criteria. These spatially resolved maps, help verify the reliability of our results by revealing whether parameters vary smoothly and align with the galaxy’s morphology. These maps also make it easier to search for modelling artifacts, such as abrupt discontinuities or parameter saturation. This enables a direct comparison of physical property distributions with features seen at other wavelengths or in simulations, providing a useful check on our assumptions. Across the sample, we find that stellar mass and SFR peaks are generally aligned. In contrast, the mass-weighted ages and FUV attenuation exhibit more complex and asymmetric patterns (see Appendix \ref{app:othersourcesCIGALE}).

\subsection{Fitting the SFR-[CII] relation}\label{sec:likelihood}

To study the relationship between the SFR obtained by the SED modelling and the [CII] luminosity, we parametrise this relation as a power law with an intrinsic scatter that accounts for the variations of the physical conditions:
\begin{equation}
\label{eq:fulllaw}
\log_{10}(\Sigma_\mathrm{{SFR}}) = \log_{10}(C) + \beta \log_{10}(\Sigma_\mathrm{[CII]}) \pm \sigma_{\rm i},
\end{equation}
where $\beta$, $C$, $\Sigma_\mathrm{[CII]}$, $\Sigma_\mathrm{SFR}$ and $\sigma_{\rm i}$ are the slope, the normalisation constant, the [CII] surface brightness, the SFR surface density, and the intrinsic scatter, respectively.

To estimate the slope and scatter of the [CII]-SFR relation, we use a Bayesian fit relying on a maximum likelihood estimator. We construct a likelihood that properly accounts for both normal distribution of [CII] maps error and log-normal nature of SFR distributions, but also estimates the intrinsic scatter. To mitigate potential degeneracies between the normalisation factor and the slope when fitting the data, we introduce a normalised [CII] surface brightness :
\begin{equation}
g_\mathrm{k,norm} = g_\mathrm{k} / g_0,
\end{equation}
where $g_\mathrm{k}$ represents the [CII] surface brightness measurement for each data point $k$, and the subscript $\rm norm$ indicate normalisation by the reference value $g_0$. The parameter \(g_0\) is a fixed normalisation pivot introduced to reduce the degeneracies between the slope \(\beta\) and the normalisation constant \(C\), and is set to the median [CII] surface brightness in our sample, ($\log_{10}(g_0) = 7.5\; \mathrm{L}_\odot\, \mathrm{kpc}^{-2}$).  The slope \(\beta\) is independent of the choice of \(g_0\), while the normalisation constant \(C\) is expressed relative to this pivot (denoted \(C_0\)). The sample $log_{10}\,g_k$ values ranges from 6.89 to 8.51, with an interquartile range of 7.23 to 7.92, confirming the representativeness of this choice.

Next, we address the statistical treatment of our data. For each data point $k$ in the [CII]-SFR plane, we consider the normalised [CII] surface brightness ($g_\mathrm{k,n}$) and SFR surface density ($s_\mathrm{k}$). To account for the uncertainties in the [CII] measurements, we model the true value $g_\mathrm{t}$ of a given $g_\mathrm{k,n}$ measurement as a normal distribution with standard deviation $\sigma_{\mathrm{g,k}}$. This uncertainty is incorporated into the likelihood through marginalisation over the $g_\mathrm{t}$ parameter.

Since the SFR estimates follow a log-normal distribution, we work with the log-quantity, denoted as $y_\mathrm{k} = \ln(s_\mathrm{k})$, which then follows a normal distribution. The mean of this distribution is given by $\mu = \ln(C_0) + \beta\ln(g_\mathrm{t})$. Finally, the standard deviation of this new distribution, $\sigma_T$, accounts for both the SFR estimate error and the intrinsic scatter of the relation: \begin{equation}
    \sigma_\mathrm{T,k} = \sqrt{\sigma_{\rm i}^2 + \sigma_\mathrm{y,k}^2}, 
\end{equation} where $\sigma_\mathrm{y,k}$ represents the SFR estimate error for a pixel $k$ computed from the $y_\mathrm{k} = \ln(s_\mathrm{k})$ distribution and $\sigma_{\rm i}$, the intrinsic scatter of the SFR-[CII] relation.

With this description, the likelihood for a given pixel is expressed as:
\begin{multline}
    p(y_\mathrm{k}, \sigma_\mathrm{y,k}, g_\mathrm{k,n},\sigma_\mathrm{g,k}|C_0,\beta,\sigma_\mathrm{i})= \\
    \int_{-\infty}^{+\infty} p(y_\mathrm{k}, \sigma_\mathrm{y,k}, g_\mathrm{t}|C_0,\beta,\sigma_\mathrm{i})p(g_\mathrm{t}|g_\mathrm{k,n},\sigma_\mathrm{g,k})\mathrm{d}g_\mathrm{t}= \\
    \frac{1}{2\pi \sigma_\mathrm{T} \sigma_\mathrm{g,k}} \int_{-\infty}^{+\infty}\exp\left(-\frac{(y_\mathrm{k}-\mu)^2}{2\sigma_\mathrm{T}^2}\right)\,\exp\left(-\frac{(g_\mathrm{k,n}-g_\mathrm{t})^2}{2\sigma_\mathrm{g,k}^2}\right)\,\mathrm{d}g_\mathrm{t}.
\end{multline}
The first term in the likelihood, $p(y_\mathrm{k}, \sigma_\mathrm{y,k}|g_\mathrm{t},C_0,\beta,\sigma_\mathrm{i})$, represents the KS relation and its associated uncertainties, while the second term, $p(g_\mathrm{t}|g_\mathrm{k,n},\sigma_\mathrm{g,k})$, accounts for the observational uncertainty on the [CII] surface brightness measurements. \\

By marginalising over $g_\mathrm{t}$, we propagate these uncertainties into our parameter estimation. The marginalisation is performed numerically by integrating on the $g_\mathrm{k,n} \pm 5(\sigma_\mathrm{g,k})$ range spanned by a grid of 1000 linearly spaced values. This covers 99.99994\% of the probability mass for a normal distribution, thus capturing nearly the entire range of plausible true [CII] surface brightness values. To estimate the model parameters, we maximise the total log-likelihood across all data points. This is equivalent to minimising the following negative log-likelihood: \begin{equation}\label{eq:Likelihood}
    \sum_k -\ln\left(p\left(y_\mathrm{k}|g_\mathrm{k,n},\sigma_\mathrm{g,k},C_0,\beta,\sigma_\mathrm{T}\right)\right).
\end{equation}
Physical constraints enforcing positivity ($C>0,~\sigma_{\rm i}>0$) were imposed via flat, uniform priors restricted to the physically meaningful domain of positive values. Parameter optimisation was performed using L-BFGS-B (from \texttt{scipy.minimize} function), a boundary-aware algorithm specifically designed for constrained optimisation.

To assess potential biases in this fitting methodology, we perform an end-to-end numerical simulation that replicates the full observational chain, from mock source generation, noise injection and resolution homogenisation to the inference of the physical relation parameters. This reveals no systematic parameter estimation biases on a simulated population similar to our sample (see Appendix\,\ref{app:NumSimBias}). However, when considering individual galaxies with a limited amount of independent data points, we obtain unreliable results highlighting the necessity of combining multiple objects to recover the proper physical relation.

\subsection{Uncertainty quantification}\label{sec:resampling}
To evaluate the robustness of our results and quantify uncertainties in the model parameters, we implement a systematic resampling strategy. This approach uses the ensemble of the 100 perturbed [CII] maps per source in our sample and their corresponding SFR estimates (derived in Sects.\,\ref{sec:ObsMocks} and \ref{sec:SEDfit}), combined with the unperturbed dataset. We investigate three distinct methods to disentangle sources of uncertainty, applying each method 1000 times to generate robust parameter distributions. The first one, called source bootstrapping, consists in selecting 13 sources with replacement of the original noise realisation. This is used to test the sample representativeness. The second one, called noise sampling, selects the original 13 sources but with a random noise realisation, enabling an assessment of the impact of observational noise. Finally, we perform a hybrid resampling by selecting 13 sources randomly as well as a random noise realisation to combine both effects. For the last two methods, we account for the additional noise introduced by the resampling process by using $\sqrt{2}$ times larger uncertainties in our fitting procedure.

In addition to our primary resampling methods, we explore alternative fitting methods. Details of these supplementary analyses, including \texttt{linmix}\footnote{Python port of \citet{Kelly2007} LINMIX\_ERR IDL code, available at \href{https://github.com/jmeyers314/linmix.git}{https://github.com/jmeyers314/linmix.git}.} Bayesian regression and Markov Chain Monte Carlo (MCMC) techniques, are presented in Appendix\,\ref{app:otherfittingmethods}. The results of these secondary approaches, are summarised in Table\,\ref{tab:fits} for direct comparison with the primary resampling methods, and are discussed in Sect.\,\ref{sec:resCIISFR}.

\begin{table}
\centering
\caption{Best-fit parameters and uncertainties for the $\Sigma_\mathrm{SFR}$-$\Sigma_\mathrm{CII}$ relation derived from the different fitting methods described in Sect.\,\ref{sec:likelihood}.}
\begin{tabular}{lccc}
\hline\hline\vspace{-0.8em}\\
Fitting Method & $\log(C_0)$\tablefootmark{a} & $\beta$ & $\sigma_{\rm i}$ \\\vspace{-1em}\\
\hline\vspace{-0.8em}\\
\multicolumn{4}{c}{Likelihood Model + L-BFGS-B integration}\\
\hline\vspace{-0.8em}\\
Original Fit & 0.30 & 0.87 & 0.19 \\\vspace{-0.8em}\\
Source Bootstrapping & $0.30^{+0.05}_{-0.04}$ & $0.86^{+0.15}_{-0.11}$ & $0.19^{+0.03}_{-0.03}$ \\\vspace{-0.8em}\\
Noise Sampling & $0.33^{+0.02}_{-0.02}$ & $0.80^{+0.05}_{-0.06}$ & $0.20^{+0.02}_{-0.02}$ \\\vspace{-0.8em}\\
Hybrid Resampling & {$0.34^{+0.05}_{-0.04}$} & {$0.80^{+0.15}_{-0.13}$} & {$0.19^{+0.04}_{-0.03}$} \\\vspace{-0.8em}\\
\hline\vspace{-0.8em}\\
\multicolumn{4}{c}{Likelihood Model + MCMC}\\
\hline\vspace{-0.8em}\\
Original Likelihood & {$0.30$} & {$0.87$} & {$0.19$} \\\vspace{-0.8em}\\
Normalised Likelihood & $0.30^{+0.03}_{-0.03}$ & {$0.87^{+0.09}_{-0.09}$} & $0.20^{+0.02}_{-0.02}$ \\\vspace{-0.8em}\\
\hline\vspace{-0.8em}\\
\multicolumn{4}{c}{Linmix}\\
\hline\vspace{-0.8em}\\
All Pixels & $0.37$ & $0.92$ & $0.18$ \\\vspace{-0.8em}\\
Region-Averages (B23) & $0.22^{+0.04}_{-0.04}$ & $1.02^{+0.12}_{-0.12}$ & $0.06^{+0.08}_{-0.08}$  \\\vspace{-0.8em}\\
\hline
\end{tabular}
\tablefoot{Parameter uncertainties, when given, are computed using the 16th, 50th, and 84th percentiles of the associated distributions.\tablefoottext{a}{We present the log($C_0$) using a normalisation factor ($g_0=10^{7.5}$) since the reported uncertainties have been derived using this $g_0$ cannot easily be extrapolated back to the un-normalised space. However, a subtraction of 7.5 from log($C_0$) allows to obtain the actual relation intercept.}}
\label{tab:fits}
\end{table}

\section{Results}\label{sec:results}
\subsection{Resolved SFR-[CII] relation}\label{sec:resCIISFR}

\begin{figure}
\centering
     \includegraphics[width=\linewidth]{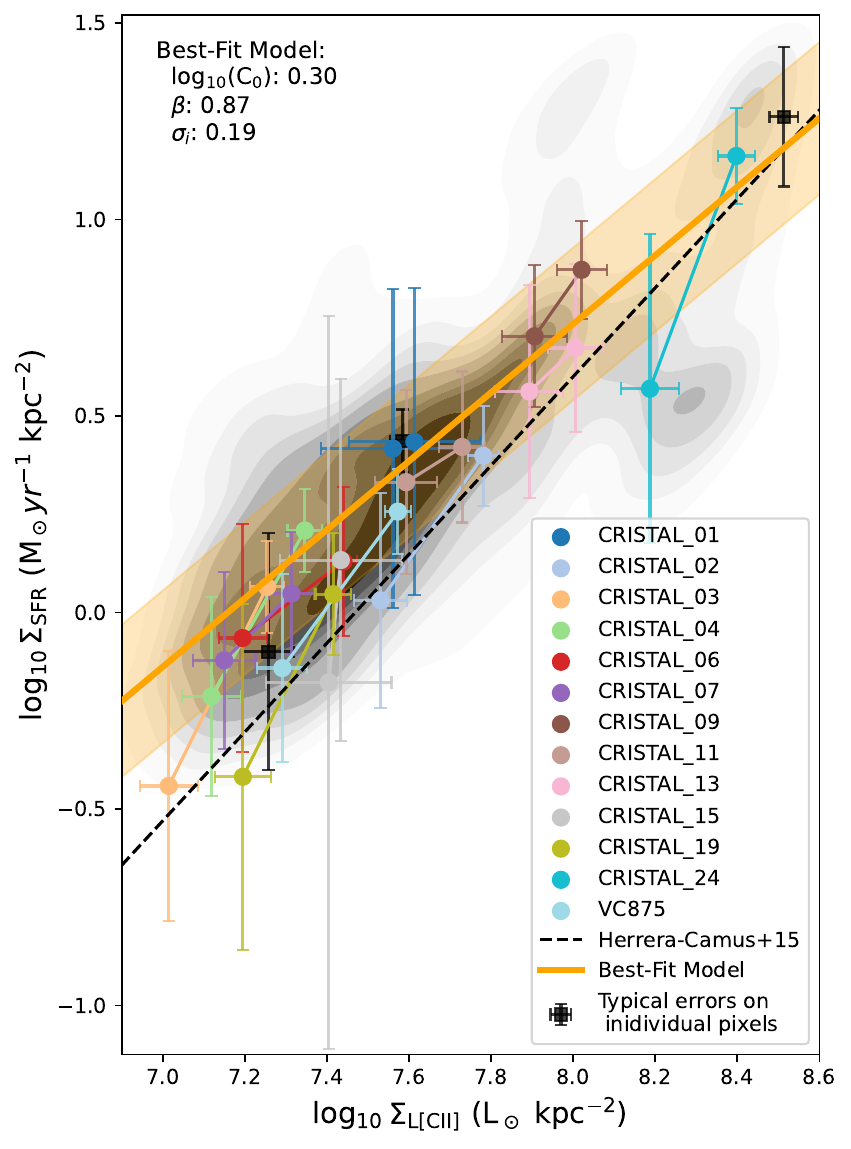}
     \caption{Resolved \([\text{CII}]\)-SFR relation for galaxies at \(z \sim 5\). The grey contours show the individual pixel distribution for the whole sample. Each pair of coloured points represents a high/low \([\text{CII}]\) density average region for each galaxy of the sample following \citetalias{BetAccGui23KS} methodology. The orange line and shaded region shows our best-fit power-law model from the tailored likelihood-based approach with the parameters in the top-left corner (where $\mathrm{log}_{10}(\mathrm{C}_0)$ is the value at $g_0=10^{7.5}\,\mathrm{L}_\odot\,\mathrm{kpc}^{-2}$). The black dashed line presents the extrapolation of the relation derived from a spatially resolved sample of 46 nearby galaxies by \citet{Herrera-Camus_2015}. We represented the typical errors bars for the individual pixels across the sample with three black squares.}
     \label{fig:LCII-SFR}
\end{figure}

\begin{figure*}
\centering
     \includegraphics[width=\textwidth]{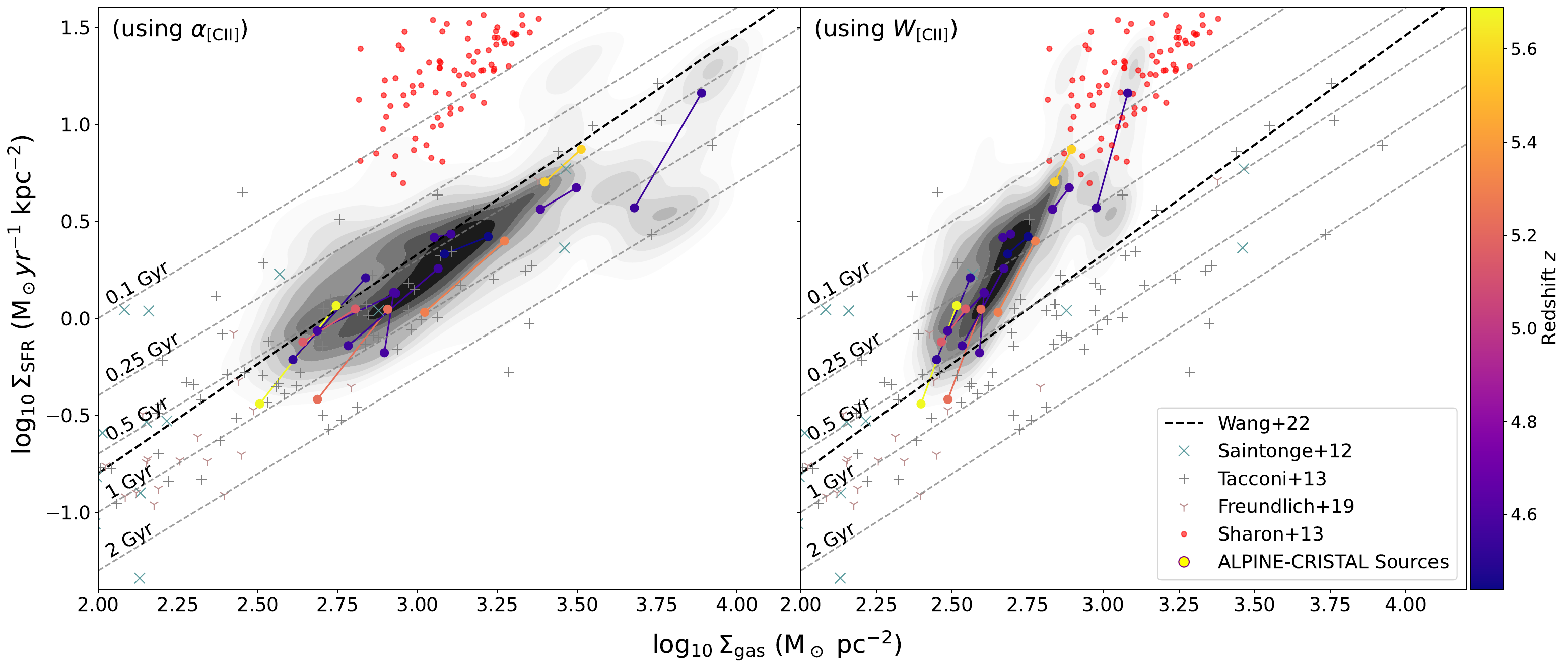}
     \caption{Kennicutt-Schmidt relation using the $\alpha_\mathrm{[CII]}$ conversion factor (left) and the $W_{\mathrm{[CII]}}$ version (right). The grey contours show the distribution of individual pixel measurements across the entire sample; the pair of points are the \citetalias{BetAccGui23KS} region-averages for individual galaxies, colour-coded by redshift. The grey dashed-lines are showing constant gas depletion time. The black dashed line represent the \cite{Wang2022} KS relation for local Universe and high redshift main sequence galaxies. We show CO measurements of the global KS relation from the low-z COLD GASS sample (\citealt{Saintonge2012}, crosses) and at redshifts up to $z\sim2.5$ from the PHIBBS (\citealt{Tacconi2013}, plus signs) and PHIBBS2 (\citealt{Freundlich2019}, three-branch stars) programs, together with a resolved $z=2.6$ lensed starburst (\citealt{Sharon_2013}, red points).}
     \label{fig:KS-SFR_compil}
\end{figure*}

We begin by fitting the resolved [CII]–SFR surface densities relation for our $z \sim 5$ galaxy sample using our tailored likelihood-based fitting routine, applied to the original noise realisation of all 13 galaxies. This yields a best-fit power-law model with $\log(C_0) = 0.30$,~$\beta = 0.87$, and $\sigma_{\rm i} = 0.19\,\text{dex}$. At this stage, we just derive the best fit, and formal uncertainties are not directly obtained. To assess the precision of these results, we compare them to those obtained with our primary resampling strategies. Using population bootstrapping, we find parameter estimates that are consistent with the baseline fit and show uniform relative uncertainties of $\sim 15\%$ for all parameters: $\log(C_0) = 0.30^{+0.05}_{-0.04}$, $\beta = 0.86^{+0.15}_{-0.12}$, and $\sigma_{\rm i} = 0.19^{+0.03}_{-0.03}\,\text{dex}$. These uncertainties remain stable across 1000 resamplings, despite the modest sample size ($N=13$), indicating that the measured relation is not dominated by individual outliers. In contrast, the noise sampling approach yields a slightly flatter slope ($\beta = 0.80^{+0.05}_{-0.06}$) and a marginally higher intrinsic scatter ($\sigma_{\rm i} = 0.20^{+0.02}_{-0.02}$). The hybrid resampling method, shows a combination of both trends ($\beta = 0.80^{+0.15}_{-0.13},~\sigma_{\rm i}=0.19^{+0.04}_{-0.03}$). The largest error bars on the slope arise from source bootstrapping, highlighting the influence of sample selection, whereas noise resampling yields lower slope values. Importantly, the intrinsic scatter remains remarkably stable across all methods, providing a reliable measure of how tightly [CII] emission traces star formation in our sample. These results are summarised in Table\,\ref{tab:fits}.

Figure \,\ref{fig:LCII-SFR} shows the relation between SFR surface density and [CII] surface brightness for our $z \sim 5$ sample. We present the best-fit model derived from our tailored likelihood-based approach for both the original sample and the region-averaged pairs following the \citetalias{BetAccGui23KS} methodology. For comparison, we also plot the relation from \citet{Herrera-Camus_2015}, which is based on spatially resolved measurements at $\sim0.2$--$1.5$\,kpc scales for 46 nearby galaxies ($d<30\,{\rm Mpc}$). While the majority of our $z \sim 5$ sample lies above their local [CII]–SFR relation, our resolved analysis yields a shallower slope and similar intrinsic scatter compared to the lower-redshift study of \citeauthor{Herrera-Camus_2015} ($\beta = 1.13 \pm 0.01$, $\sigma_{\rm i} = 0.21$\,dex, reported without uncertainty). In addition to the redshift difference, the parameter space covered by the two samples is very different: the \citet{Herrera-Camus_2015} sample spans $4.4 < \log \Sigma_{\rm [CII]}\,(\mathrm{L_\odot\,kpc^{-2}}) < 6.5$, whereas our sample covers $6.8 < \log \Sigma_{\rm [CII]} < 8.6$. This difference likely contributes to the observed slope discrepancy as it may reflect the evolution in ISM physical conditions across cosmic time, such as increased turbulence, lower metallicity, or changes in ISM structure at high redshift, yielding different dynamical range in $\Sigma_{\rm [CII]}$ as well as $\Sigma_{\rm SFR}$. We emphasise that methodological differences also play a critical role, especially on observation-based studies. For instance, \citet{Herrera-Camus_2015} employed a least-squares bisector, which neglects pixel covariance. In contrast, our tailored approach explicitly accounts for pixel covariance and highlights the sensitivity of the slope estimate to this effect, as demonstrated by the various tests conducted below.

We also explore alternative fitting methods to further validate our findings, as presented in Appendix\,\ref{app:otherfittingmethods}. MCMC analysis of the unperturbed dataset reproduces the best-fit parameters of the baseline fit, but initially yields unrealistically narrow error bars. For example, the MCMC approach yields an intrinsic scatter of $\sigma_{\rm i}=0.19\pm0.002$, showing error bars an order of magnitude smaller than uncertainties obtained via hybrid ressampling ($\sim 0.03$). This discrepancy likely arises from neglected correlations in the data. To correct for this, we use the likelihood normalised by half the number of pixels per beam, resulting in more realistic uncertainties that closely match those from the hybrid resampling approach (e.g. $\sigma_{\rm i}=0.20\pm0.02$). Using the \texttt{linmix} Bayesian regression method on the original dataset produces a slightly steeper slope ($\{\beta = 0.92\pm0.02$) and comparable intrinsic scatter ($\sigma_{\rm i} = 0.18\pm0.04\,\text{dex}$), again with very narrow uncertainties. To mitigate the impact of pixel covariance, we also apply \texttt{linmix} to region-averaged points following the \citetalias{BetAccGui23KS} methodology that consists in averaging the SFR and $\Sigma_\mathrm{[CII]}$ measurements over large regions to obtain a sufficiently elevated signal-to-noise ratio and by construction gives a pair of points per galaxy corresponding to the low and high regime of [CII] surface brightness (pairs of points in Fig.\,\ref{fig:LCII-SFR} and Fig.\,\ref{fig:KS-SFR_compil}). This approach yields a steeper slope than the pixel-based fit ($\beta = 1.02 \pm 0.12$) but a smaller and poorly constrained scatter ($\sigma_{\rm i} = 0.06 \pm 0.08\,\text{dex}$).

The MCMC and resampling methods yield parameter estimates that agree within $\sim1\sigma$, demonstrating the robustness of our resolved analysis. In contrast, the \texttt{linmix} method yields a slightly higher slope and a poorly constrained intrinsic scatter, indicating a mild tension with the other primary approaches. These comparisons highlight the importance of properly accounting for pixel covariance, as methods that neglect this effect can underestimate uncertainties or fail to reliably constrain the intrinsic scatter.

Because of the limitations of our model when applied on a per-galaxy basis, that is with restricted $\Sigma_{\rm [CII]}$ dynamic ranges (see Sect.\,\ref{sec:likelihood}), we are unable to recover resolved [CII]–SFR relations per galaxy. We therefore adopt the \texttt{linmix} method for simplicity, acknowledging that it does not properly account for pixel correlations. This method yields an average slope of 1.27, which is consistent with the mean value of 1.21 derived from \citet{Juno} using spatially bin-averaged measurements on the same ALPINE-CRISTAL sample. As this value is manually measured from their Fig.\,12, no formal uncertainty is provided. This agreement suggests that, despite the methodological differences and the limited dynamic range within single galaxies, the underlying scaling relation inferred from spatially resolved data remains broadly compatible with results based on coarser spatial binning.

Our results offer a valuable comparison to the theoretical predictions of \cite{Ferrara19CII}, who used spatially resolved models to investigate the [CII]-SFR relation within galaxies. Their work explored how local ISM conditions (e.g. density, metallicity) shape the resolved $\rm \Sigma_{[CII]}-\Sigma_{SFR}$ relation, finding that it steepens at high surface densities due to $\rm \Sigma_{[CII]}$ saturation. We
observe a similar trend on local scales, with the mean slope within individual galaxies aligning with their predictions. However, our relation is significantly shallower for the population as a whole, confirming that this steepening is a local effect. This contrast highlights the critical role of spatial scale in interpreting the [CII]-SFR relation. Furthermore, it underscores the need for both local and population-wide perspectives to fully understand the diverse ISM conditions at high redshift.

\subsection{[CII]-to-gas conversion factors}\label{sec:CIIconversion}

To estimate the gas mass, we first use the prescription of \citet{ZanellaCII}: the conversion factor is constant with a value of $\alpha_{\rm [CII]}=31\,\mathrm{M}_\sun\,\mathrm{L}_\sun^{-1}$, and the gas mass surface density is derived as:
\begin{equation}\label{eq:zanella}
\Sigma_\mathrm{gas} = \alpha_\mathrm{[CII]}\,\Sigma_\mathrm{[CII]}.
\end{equation}

It is well established that [CII] emission can originate from multiple ISM phases, primarily PDRs and molecular gas, yet the contributions of these phases to the total [CII] luminosity remain debated and are sensitive to local ISM conditions. Consequently, the dominant gas phase traced by [CII] can vary across different environments, particularly at high redshift where ISM properties often diverge from those in local galaxies. This diversity has a direct impact on the appropriate [CII]-to-gas conversion factor \citep[e.g.][]{Madden20CII, Vizgan_2022}. Recent theoretical work reinforces this perspective: \citet{vallini2025}, using galaxies from the SERRA cosmological zoom-in simulation \citep{Pallottini2022}, derive a relation varying with the [CII] surface brightness:
\begin{equation}\label{eq:livia}
\log W_{\mathrm{[CII]}} = -0.506 \, \log \Sigma_\mathrm{[CII]} + 4.933,
\end{equation}
where the $W_\mathrm{[CII]}$ notation is used to express the resolved, surface brightness-dependent conversion factor, in contrast to the global $\alpha_{\rm [CII]}$ value. This relation exhibits a typical 1$\sigma$ dispersion of 0.4\,dex. The dependence on $\Sigma_\mathrm{[CII]}$ encapsulates the expectation that brighter [CII] regions are denser and more metal enriched, two conditions that drive higher [CII] emission efficiency per unit gas mass. Across the $\rm 6.8<\log \Sigma_{[CII]}\,(L_\sun\,kpc^{-2})<8.6$ range probed in our sample, $W_\mathrm{[CII]}$ varies from $\rm 31\,M_\sun\,L_\sun^{-1}$ in the faintest regions down to $\rm 3.8\,M_\sun\,L_\sun^{-1}$ in the brightest regions. This prescription can therefore lead to a steeper slope in the KS relation derived from [CII], particularly at high surface densities. It is important to note that the conversion from \citet{vallini2025} yields a relation computed for the total cold gas content. This total content encompasses both atomic and molecular phases, thereby accounting for the outer layers of PDRs and the fully molecular clumps within clouds.

To explore the implications of this conversion factor on our interpretation of [CII]-based observations, we include it in our analysis. Using the $W_\mathrm{[CII]}$ formula from Eq.\,\eqref{eq:livia} in place of $\alpha_\mathrm{[CII]}$ in Eq.\,\eqref{eq:zanella} yields:
\begin{equation}
\Sigma_{\mathrm{[CII]}} \propto \Sigma_{\mathrm{gas}}^{1/0.494}.
\end{equation}
Substituting this relation into Eq.\ref{eq:fulllaw} gives:
\begin{equation}\label{eq:sfrWcii}
\Sigma_{\mathrm{SFR}} \propto \Sigma^\beta_\mathrm{[CII]} \textrm{ and thus } \Sigma_{\mathrm{SFR}} \propto \Sigma_{\mathrm{gas}}^{\beta/0.494}.
\end{equation}
This implies that for a given slope $\beta$ of the SFR–[CII] relation, the slope of the SFR–gas relation is expected to be approximately twice as steep. In contrast, the constant $\alpha_{\rm [CII]}$ prescription leaves the slope $\beta$ unaffected.

\subsection{Impact on the Kennicutt–Schmidt relation}\label{sec:KSrelation}

Figure\,\ref{fig:KS-SFR_compil} presents the resolved KS relation for our $z\sim5$ sample, contrasting two [CII]-to-gas conversion prescriptions. The left panel adopts the constant $\alpha_{\rm [CII]}$ from \citet{ZanellaCII}, and the right applies the surface brightness-dependent $W_{\rm [CII]}$ from \citet{vallini2025}. In both cases, we overlay a set of reference data from CO-based global KS relations from the low-redshift COLD GASS sample (\citealt{Saintonge2012}), as well as higher-redshift measurements from PHIBBS (\citealt{Tacconi2013}) and PHIBBS2 (\citealt{Freundlich2019}). The resolved $z=2.6$ lensed-starburst of \citet{Sharon_2013} is also shown, providing a benchmark for extreme star-forming environments. The black dashed line in both panels represents the KS relation for local and high-redshift main-sequence galaxies from \citet{Wang2022}. Due to the substantial dispersion inherent to both conversion factor prescriptions, our fitting methodology cannot robustly constrain the intrinsic scatter of the relation. As such, the following discussion provides a qualitative assessment of the observed trends, and reported parameter values should be interpreted with caution.

When adopting a constant $\alpha_{\rm [CII]}$, our sample is compatible with the \citet{Wang2022} results, despite small slope difference ($\beta=1.13\pm0.09$ for \citealt{Wang2022}, $\beta=0.87\pm0.06$ in this work). The data align with constant gas depletion times of $\sim0.5-1$\,Gyr, extending the locus of the low- and intermediate-redshift CO-based studies (\citealt{Saintonge2012}; \citealt{Tacconi2013}; \citealt{Freundlich2019}) to higher gas and SFR surface densities. Interestingly, these are $\sim2-3$ times shorter than those usually found for local star-forming spirals at the similar physical scales using CO-based $\rm H_2$ tracers \citep[e.g][]{Schruba2011, Leroy2023, Villanueva2021, Villanueva2024a}. Within individual galaxies, we observe moderate internal variations, with denser regions showing shorter depletion times, but no evidence for extreme starburst-like behaviour with depletion times above 0.25\,Gyr throughout.

Applying the $W_{\rm [CII]}$ prescription fundamentally alters the picture. The derived slope steepens to $\beta=1.75\pm0.14$, and the sample bridges the gap between the low-redshift CO-based locus and the regime occupied by the $z=2.6$ starburst of \citet{Sharon_2013}. Several regions within our galaxies, particularly in CRISTAL\_09, CRISTAL\_13, and CRISTAL\_24, overlap with these resolved starburst measurements. This change is accompanied by a pronounced decrease in depletion times across the most [CII]-emitting regions, with values reaching as low as 0.1\,Gyr, characteristic of starburst systems, in the brightest [CII] regions. Importantly, this shift towards shorter depletion times is not restricted to isolated, compact star-forming regions, but is observed globally across the galaxies. While increasing spatial resolution is generally expected to broaden the dynamic range of depletion times and reveal localised starburst-like environments, we do not observe such a trend here. Instead, the $W_{\rm [CII]}$ conversion produces a systematic reduction in depletion times throughout all regions, compressing the inferred gas mass range. This raises the possibility that, under certain conversion prescriptions, main-sequence galaxies could exhibit global properties reminiscent of starbursts, challenging the classical separation between main-sequence and starburst systems.

Recent results from the THESAN-ZOOM simulations \citep{Shen2025} show a similar trend towards shorter depletion timescales for $\Sigma_{\rm gas} > 100\,M_\sun\,\mathrm{pc}^{-2}$. In their sample of galaxies with stellar masses up to $10^{11}\,M_\odot$ at $z\sim3$ and $10^{9.6}\,M_\odot$ at $z\sim5$, resolved depletion times for neutral gas (HI+H$_2$) can reach as low as $0.1$ Gyr in the densest regions (see their Fig.\,8). This is consistent with the short depletion times inferred in our analysis.

Additionally, we colour-code our sample by redshift in Figure\,\ref{fig:KS-SFR_compil} and find no clear redshift dependence of the relative position of the galaxies within the KS plane. 

Under the constant $\rm \alpha_{[CII]}=31\,M_\sun\,L_\sun^{-1}$, high-redshift main-sequence galaxies exhibit KS behaviour remarkably similar to their low-redshift counterparts, as traced by CO. In contrast, the $W_{\rm [CII]}$ approach reveals a population of short-depletion-time regions that bridge the loci of main-sequence galaxies and starburst regimes, demonstrating the strong sensitivity of inferred gas properties, and thus depletion time estimates, to the adopted [CII]-to-gas calibration. As a result, our ability to confidently study the KS relation at high redshift is compromised.

\section{Discussion}\label{sec:discussion}

\begin{figure}
\centering
     \includegraphics[width=\linewidth]{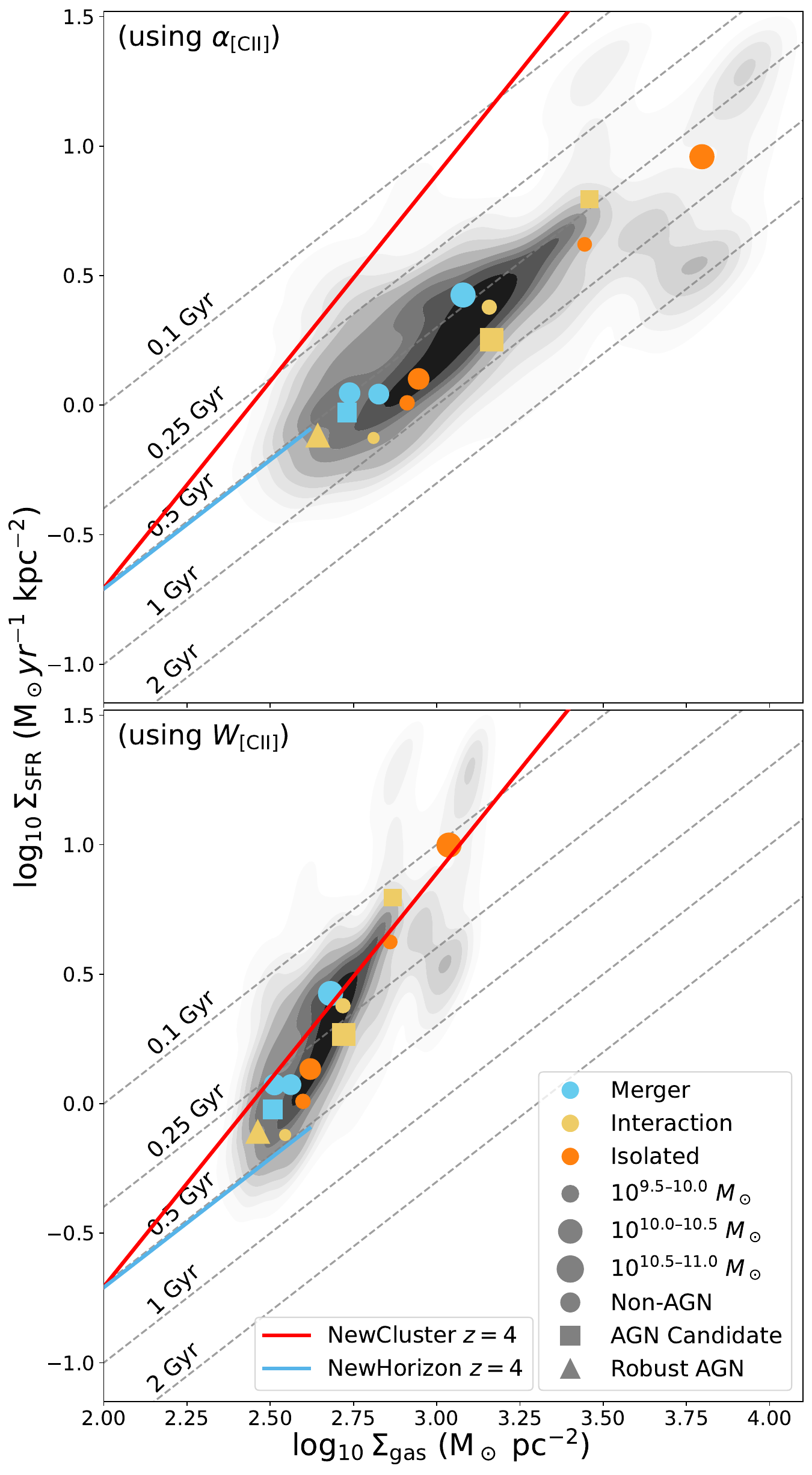}
     \caption{Similar to Fig.\,\ref{fig:KS-SFR_compil}. This time, symbols indicate integrated measurements for individual galaxies, colour-coded by interaction state: light blue for major mergers, light yellow for interacting systems, and orange for isolated galaxies, following the classification of \citet{herreracamus2025almacristalsurveygasdust}. Symbol size encodes the total stellar mass and symbol shape denotes AGN identification: circles for non-AGN, squares for AGN candidates, and triangles for robust AGN as per Ren et al. (in prep). The solid blue line shows the $z=4$ relation from \citet{Kraljic24}, based on the 25\% most massive galaxies in the NewHorizon simulation, while the solid red line indicates the NewCluster analogue (Yi et al. in prep). Both simulation relations are plotted over their effective gas mass ranges for direct comparison with our sample.}
     \label{fig:KS-Discussion}
\end{figure}

To further investigate the possible physical drivers of the derived KS relations, we construct a diagnostic plot shown in Fig.\,\ref{fig:KS-Discussion} encoding AGN activity, stellar mass, and merger status for our sample by changing symbol type, size, and colour respectively. AGN identification follows Ren et al. (in prep), while merger status is classified according to \citet{herreracamus2025almacristalsurveygasdust}. Additionally, environmental classification distinguishes between field galaxies and those residing in the $z\sim4.57$ protocluster structure (CRISTAL\_13, CRISTAL\_15, and VC875; see \citealt{Lemaux2018,Staab2024}). Visual inspection of the KS diagram reveals no clear systematic offset or trend associated with AGN presence, merger stage, stellar mass, or environment across the KS plane. Within the dynamic range probed by our study, none of these properties seems to have a significant impact on the star formation. Nevertheless, we caution that the uncertainties and limited sample size may reduce sensitivity to subtle or systematic effects.

We also show KS relations for the 25\% most massive galaxies at $z=4$ in both the NewHorizon \citep{DuboisNH,Kraljic24} and NewCluster (Yi et al., in prep.) simulations, presented in their respective effective gas surface density regimes. The main difference between the two simulations is environmental: NewCluster targets a high-density proto-cluster region, while NewHorizon samples an average-density field. Both use similar physical models for star formation and feedback. In the constant $\alpha_{\rm [CII]}$ case, our observed sample aligns well with predictions from the New Horizon simulation, though this simulation primarily probes lower stellar masses ($10^{8.0-9.5}\,\mathrm{M}_\sun$) than our galaxies ($10^{9.5-10.6}\,\mathrm{M}_\sun$). In comparison, the 25\% most massive galaxies in the NewCluster simulation span a similar stellar mass range ($M_* = 10^{9.5-10.5}\,M_\sun$). We find that employing a density-dependent $W_{\rm [CII]}$ conversion factor aligns with their results. This compatibility could support the fact that the stellar mass affects the slope of the KS relation for high-redshift galaxies, specifically a steepening of the slope as the total stellar mass of the population increases, as hinted by the $z=2-3$ relations presented in \citet{Kraljic24}. 

The contrasting behaviours obtained assuming the two different gas-conversion prescriptions, and their respective alignment with different simulation, show the need for reliable [CII]-to-gas mass conversion methods that account for the ISM conditions prevailing at a given redshift and in a specific population of galaxies. Even though we cannot fully exclude it at this stage, the lack of correlation with AGN or merger status further points to intrinsic ISM properties, such as density, turbulence, and metallicity, as the dominant factors shaping the [CII]-to-gas calibration and, by extension, the inferred star formation law.

These results highlight the limitations of adopting a universal [CII]-to-gas conversion factor and motivate the development of physically motivated, simulation-calibrated or locally-calibrated prescriptions that account for galaxy mass, environment, and ISM state. Joint analysis of resolved observations and simulations, as demonstrated here, is essential for building an accurate picture of the star formation mechanisms in the high-redshift Universe.

\section{Conclusions}\label{sec:conclusion}

We present a spatially resolved analysis of the [CII]–SFR and Kennicutt–Schmidt relations in a sample of 13 main-sequence galaxies at redshifts $4 < z < 6$, combining ALMA, JWST, and HST data. Our pixel-by-pixel SED modelling approach applied to resolution-homogenised multi-wavelength datasets allows us to probe the physical conditions of star formation down to correlated kiloparsec scales in the early Universe.

We develop a new comprehensive statistical framework, which explicitly accounts for correlated noise and pixel covariance introduced by resolution homogenisation. By combining noise sampling, source bootstrapping, and hybrid resampling techniques, we reliably estimate the intrinsic scatter and slope of the spatially resolved [CII]-SFR relations. Comparison with classical fitting methods demonstrates that neglecting pixel correlations leads to underestimated uncertainties. This shows the relevance of our approach for robust parameter inference.

Our analysis reveals that the resolved [CII]–SFR relation exhibits a slope of $0.87 \pm 0.15$ and an intrinsic scatter of $0.19 \pm 0.03$\,dex. These values are both shallower and tighter than those reported in many studies of the local Universe, suggesting a possible saturation of the SFR emission at high [CII] surface densities, population wise, while confirming a strong correlation between [CII] and SFR at $z \sim 5$. This tight relation supports the use of [CII] as a reliable tracer of star formation in early main-sequence galaxies.

In contrast, the resolved KS relation is highly sensitive to the assumption of the [CII]-to-gas conversion factor. Using a constant $\alpha_{\rm [CII]}$ \citep{ZanellaCII} yields a slope of $0.87$ and depletion timescales of roughly 0.5 to 1\,Gyr, consistent with expectations for main-sequence galaxies. However, adopting the $\Sigma_{\rm [CII]}$-dependent $W_{\rm [CII]}$ prescription \citep{vallini2025} results in a much steeper slope of $1.75$. The regions with our lowest probed gas surface densities still behave as the disc sequence with depletion timescales of the order of 0.5\,Gyr, but high-density regions exhibit shorter depletion times ($<0.1$\,Gyr), placing them in the starburst regime.

Notably, these two scenarios bracket the predictions from recent simulations: the constant $\alpha_{\rm [CII]}$ case is in broad agreement with NewHorizon results at $z=4$ and low mass ($10^{8.0-9.5}\,\mathrm{M}_\sun$), while the $W_{\rm [CII]}$ prescription produces KS slopes and depletion times compatible with the galaxies seen in NewCluster in a similar mass regime as our sample ($10^{9.5-10.5}\,M_\odot$). The range of predicted depletion times produced by these two conversion factors points to a fundamental limitation in current gas mass estimates from [CII] and illustrates the need for physically motivated conversion factors that capture the evolving ISM conditions, metallicity, and dynamical processes at high redshift.

\begin{acknowledgements} Cédric Accard acknowledges that this work of the Interdisciplinary Thematic Institute IRMIA++, as part of the ITI 2021-2028 program of the University of Strasbourg, CNRS and Inserm, was supported by IdEx Unistra (ANR-10-IDEX-0002), and by SFRI-STRAT’US project (ANR-20-SFRI-0012) under the framework of the French Investments for the Future Program. This work was supported by the « action thématique » Cosmology-Galaxies (ATCG) of the CNRS/INSU PN Astro. This work was supported by the French government through the France 2030 investment plan managed by the National Research Agency (ANR), as part of the Initiative of Excellence of Université Côte d’Azur under reference number ANR-15-IDEX-01. 
M.Bo. acknowledges support from the ANID BASAL project FB210003.
L.V. acknowledges support from the INAF Minigrant "RISE: Resolving the ISM and Star formation in the Epoch of Reionization" (PI: Vallini, Ob. Fu. 1.05.24.07.01).
E.d.C. acknowledges support from the Australian Research Council through project DP240100589.
M.A. is supported by FONDECYT grant number 1252054, and gratefully acknowledges support from ANID Basal Project FB210003 and ANID MILENIO NCN2024\_112.
A.F. is partly supported by the ERC Advanced Grant INTERSTELLAR H2020/740120, and by grant NSF PHY-2309135 to the Kavli Institute for Theoretical Physics.
R.H.-C. thanks the Max Planck Society for support under the Partner Group project "The Baryon Cycle in Galaxies" between the Max Planck for Extraterrestrial Physics and the Universidad de Concepción. R.H-C. also gratefully acknowledge financial support from ANID - MILENIO - NCN2024\_112 and ANID BASAL FB210003.
J.M. gratefully acknowledges support from ANID MILENIO NCN2024\_112.
A.N., P.S. acknowledge support from the Narodowe Centrum Nauki (NCN), Poland, through the SONATA BIS grant UMO-2020/38/E/ST9/00077.
M.P. acknowledges financial support from the project "LEGO – Reconstructing the building blocks of the Galaxy by chemical tagging" granted by the Italian MUR through contract PRIN2022LLP8TK\_001.
M.Re. acknowledges support from project PID2023-150178NB-I00 financed by MCIU/AEI/10.13039/501100011033.
V.V. acknowledges support from the ANID BASAL project FB210003 and from ANID - MILENIO - NCN2024\_112.
W.W. acknowledges the grant support through JWST programs. Support for programs JWST-GO-03045 and JWST-GO-03950 were provided by NASA through a grant from the Space Telescope Science Institute, which is operated by the Association of Universities for Research in Astronomy, Inc., under NASA contract NAS 5-03127.
S.K.Y. acknowledges support from the Korean National Research Foundation (RS-2025-00514475 and RS-2022-NR070872). 
This paper makes use of the following ALMA data: ADS/JAO.ALMA\#2017.1.00428L, ADS/JAO.ALMA\#2019.1.00226.S, ADS/JAO.ALMA\#2021.1.00280.L, ADS/JAO.ALMA\#2022.1.01118.S. 
ALMA is a partnership of ESO (representing its member states), NSF (USA) and NINS (Japan), together with NRC (Canada), MOST and ASIAA (Taiwan), and KASI (Republic of Korea), in cooperation with the Republic of Chile. The Joint ALMA Observatory is operated by ESO, AUI/NRAO and NAOJ. 
NewCluster was granted access to the HPC resources of KISTI under the allocations KSC-2022-CRE-0344 and KSC-2023-CRE-0343, and of GENCI under the allocation A0150414625. We thank Diana Ismail for sharing her data compilation of NewCluster. We acknowledge the use of Astropy \citep{Astropy}, Matplotlib \citep{Matplotlib}, Scipy \citep{Scipy}.
\end{acknowledgements}

\bibliographystyle{aa}

\begin{appendix}
\onecolumn
\section{Homogenised data}\label{app:HomogenisedData}

\begin{figure*}[h!]
\centering
   \includegraphics[width=16cm]{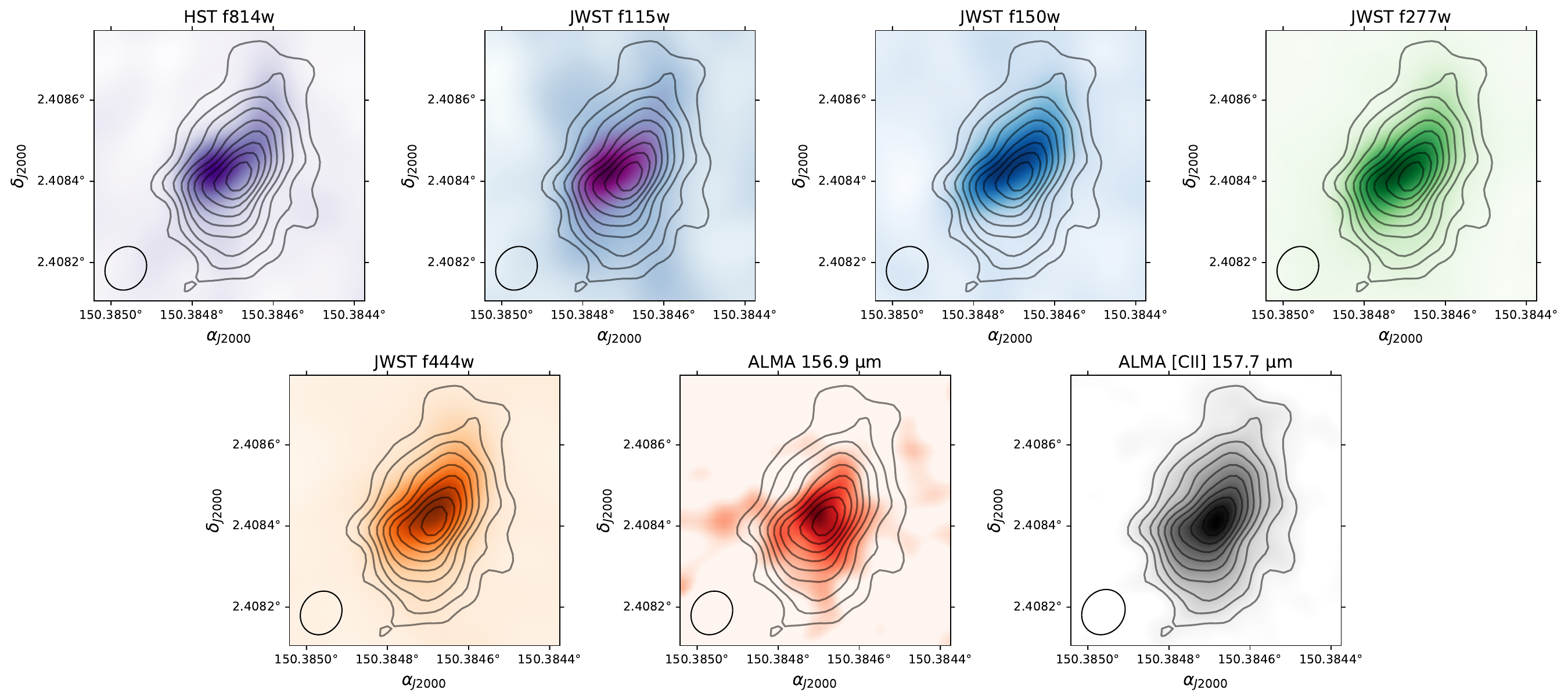}
     \caption{Resolution homogenised version of Fig.\ref{fig:VC875_native}, with the ALMA resolution as the target. The first row presents, from left to right, HST/ACS F814W (rest-frame 145\,nm), JWST/NIRCam F115W (207\,nm), F150W (270\,nm), and F277W (499\,nm). The second row shows JWST/F444W (800\,nm), ALMA dust continuum emission around $156.9\,{\rm \mu m}$, and ALMA [CII] $158\,{\rm \mu m}$ emission. The black contours are the (3+2k)$\sigma$ levels ($k\geq0$) of the [CII] luminosity map, while the ellipses in the bottom-left corners all panels indicate the synthesised ALMA beam. These images illustrate the filter coverage available across all galaxies in our sample.}
     \label{fig:VC875smoothed}
\end{figure*}

The resolution-homogenised maps of VC875 is shown in Fig.\,\ref{fig:VC875smoothed} as an example.

\begin{table*}[h!]
\section{SED modelling parameters}
    \caption{Parameter values for different modules used for the CIGALE runs.}
    \label{tab:Param}
    \centering
    \begin{tabular}{lll}
        \hline\hline\vspace{-0.8em}\\
        {Module} & {Parameter} & {Value} \\\vspace{-1em}\\
        \hline\vspace{-0.8em}\\
        \texttt{sfhdelayedbq}  & age\_main\,(Myr) & (Universe age at source redshift) - 250\,Myr\\
        & tau\_main\,(Myr) & 100, 300, 500, 700, 900, 1100, 1300 \\
        & age\_bq\,(Myr) & 0, 5, 10, 25, 50, 100, 150, 200, 250, 300, 350, 400, 450\\\rule{0pt}{2ex}
        & {r\_sfr} & 33 log-spaced values from 0 to 900.0 \\\vspace{-1em}\\
        \hline\vspace{-0.8em}\\
        \texttt{bc03} & imf & 1 \citep{Chabrier}\\
        & metallicity & 0.008 \\\vspace{-1em}\\
        \hline\vspace{-0.8em}\\
        \texttt{nebular} & logU  & $-2.0$ \\\vspace{-1em}\\
        \hline\vspace{-0.8em}\\
        \texttt{dustatt\_modified\_CF00} & Av\_ISM & 101 linearly-spaced values from 0.3 to 5 \\
        & mu & 0.5 \\
        & slope\_ISM & $-0.7$ \\
        & slpoe\_BC & $-0.7$ \\\vspace{-1em}\\
        \hline\vspace{-0.8em}\\
        \texttt{dustatt\_calzleit} & E\_BVs\_young & 100 linearly-spaced values from 0.005 to 0.500 \\
        (tested but not used) & powerlaw\_slope & 0 or [-2.0, 0.5]\\
        \hline\vspace{-0.8em}\\
        \texttt{redshifting} & redshift  & [CII] spectroscopic redshift from ALPINE \citep{bethermin2020alpine} \\\vspace{-1em}\\
        \hline
    \end{tabular}
    \tablefoot{The values for \texttt{dustatt\_modified\_CF00} are from \cite{BuatParam} whereas the ones for \texttt{dustatt\_calzleit} come from \cite{BoquienParam}. Any parameter of a given module not shown here or not discussed in Sect.\,\ref{sec:SEDfit} has been left to the default value.
    }
\end{table*}
\twocolumn

\section{Assessing the validity of the SED modelling}\label{App:SFRvsSFR}

\begin{figure}[h!]
\centering
\includegraphics[width=0.72\linewidth]{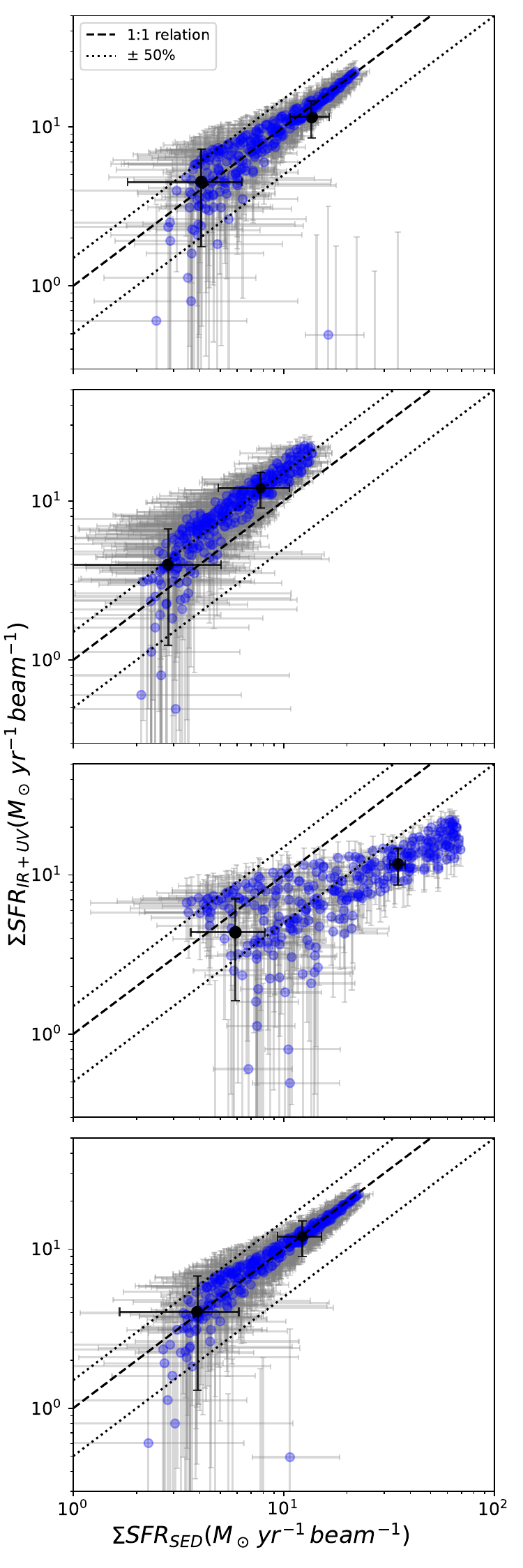}
\caption{Comparison between SFR derived from SED modelling (SFR$\mathrm{SED}$) and those measured from IR+UV observations (SFR$\mathrm{IR+UV}$) for VC875, shown for different model parametrisations. From top to bottom: (first) the fiducial model using the \citet{CF00} dust attenuation law and a delayed SFH with a recent variation episode; (second) the same, but without the recent variation episode; (third) the fiducial model without ALMA dust luminosity constraints; (fourth) the \citet{Calz1994} attenuation law model. Blue points represent individual pixels, with error bars corresponding to the 16th and 84th percentiles of the SED-derived SFR distribution (see Sect.\,\ref{sec:SEDfit}). Red arrows indicate upper limits. Black points show region-averaged SFR values computed using the \citetalias{BetAccGui23KS} method. The dashed line indicates the one-to-one relation, and dotted lines mark the $\pm$50\% range.}
\label{fig:VC875SFRcomp}
\end{figure}

We assess the impact of different parametrisations in the CIGALE SED modelling process on SFR estimates. To evaluate the reliability of each model configuration, we compare the SFR values derived by CIGALE (SFR$_\mathrm{SED}$) with those independently inferred from combined UV and far-infrared (FIR) measurements, following the methodology of \citet{BetAccGui23KS}. Figure\,\ref{fig:VC875SFRcomp} presents this comparison for VC875 as an example with each panel corresponding to a different set of CIGALE parameters. The one-to-one relation (dashed line) and the $\pm$50\% deviation lines (dotted) are included to facilitate a direct visual assessment of model performance.

It is important to emphasise that agreement between SED-based and UV+FIR SFR estimates does not necessarily guarantee the absence of systematic uncertainties or biases, since both methods may be subject to common assumptions regarding dust attenuation laws, star formation histories (SFHs), or initial mass functions (IMFs).

The top panel shows results from our fiducial model, which employs the \citet{CF00} dust attenuation law combined with a delayed SFH including a recent episode of star formation variation. This configuration achieves the closest agreement between SFR estimators, with most data points clustering near the one-to-one line.

When the recent SFH variation is excluded (second panel), the SED-based SFR estimates are systematically lower, with the majority of pixels lying above the one-to-one relation, showing a deterioration in fit quality.

Excluding the dust luminosity constraint from the modelling (third panel) results in systematic overestimation of SED-based SFRs relative to the UV+FIR values, with many points falling well beyond the $-50\%$ deviation line. In this situation, the error bars increase significantly, reflecting greater uncertainties in the fits. This highlights the critical role of including infrared constraints to properly model dust attenuation and derive reliable SFRs.

\begin{figure}[h!]
\centering
\includegraphics[width=0.7\linewidth]{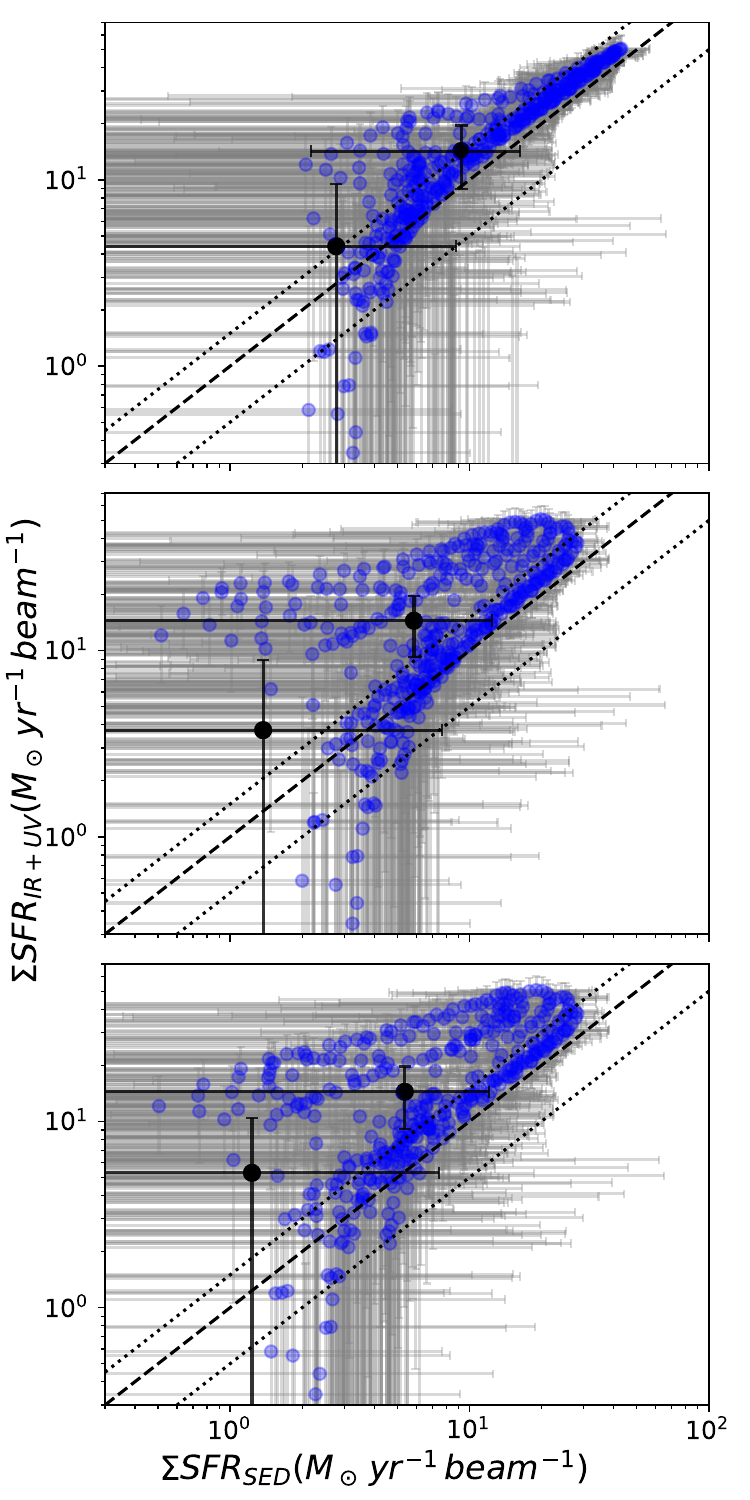}
\caption{Same as Fig.\,\ref{fig:VC875SFRcomp}, but for CRISTAL\_24. From top to bottom: Results using the \citet{CF00} dust attenuation model, results using the modified \citet{Calz1994} model with $\delta=0$, same but letting the $\delta$ parameter vary between -2.0 and 0.5}
\label{fig:C24SFRcomp}
\end{figure}

The fourth panel presents results obtained using the \citet{Calz1994} dust attenuation law. Across the sample, differences relative to the fiducial model are minimal. However, for CRISTAL\_24 (see Fig.\,\ref{fig:C24SFRcomp}), we observe the largest discrepancy between attenuation prescriptions: the \citet{CF00} law (top panel) achieves the best agreement between SFR estimators, with most data points lying close to the one-to-one line, while the fiducial model (centre panel) systematically underestimates SFRs and yields a lower fraction of well-fitted pixels. We verified that this difference is not attributable to the choice of the $\delta$ parameter in the Calzetti law, as letting $\delta$ vary produces identical results (bottom panel). Given the improved consistency for CRISTAL\_24 and the more physically motivated nature of the \citet{CF00} law, we adopt it as our default attenuation model for the entire sample.

\newpage

\newpage

\section{Validation of the likelihood model}\label{app:NumSimBias}

To evaluate the accuracy and robustness of our fitting methodology, we perform an end-to-end simulation designed to replicate the observational conditions and intrinsic properties of our dataset. Specifically, we generate 1000 independent mock realisations. For each of them, we create first gas surface density maps composed of a random number of elliptical Gaussian clumps with varying amplitudes and positions. We then convolve this map with the ALMA beam. Using a fixed star formation relation characterised by a known slope, intercept, and intrinsic scatter, we derive the corresponding SFR surface density maps on a pixel-by-pixel basis. We then simulated the intrinsic scatter by multiplying this map by a map containing log-normal fluctuations at the scale of the ALMA beam with dispersion corresponding to the intrinsic scatter and a mean value of unity. This step is important, since we measure the intrinsic scatter at the scale of the ALMA beam, and applying a pixel-based scatter would lead to a smaller scatter after smoothing the maps by the beam.

\newpage
\begin{figure}[h!]
\centering
\includegraphics[width=0.9\linewidth]{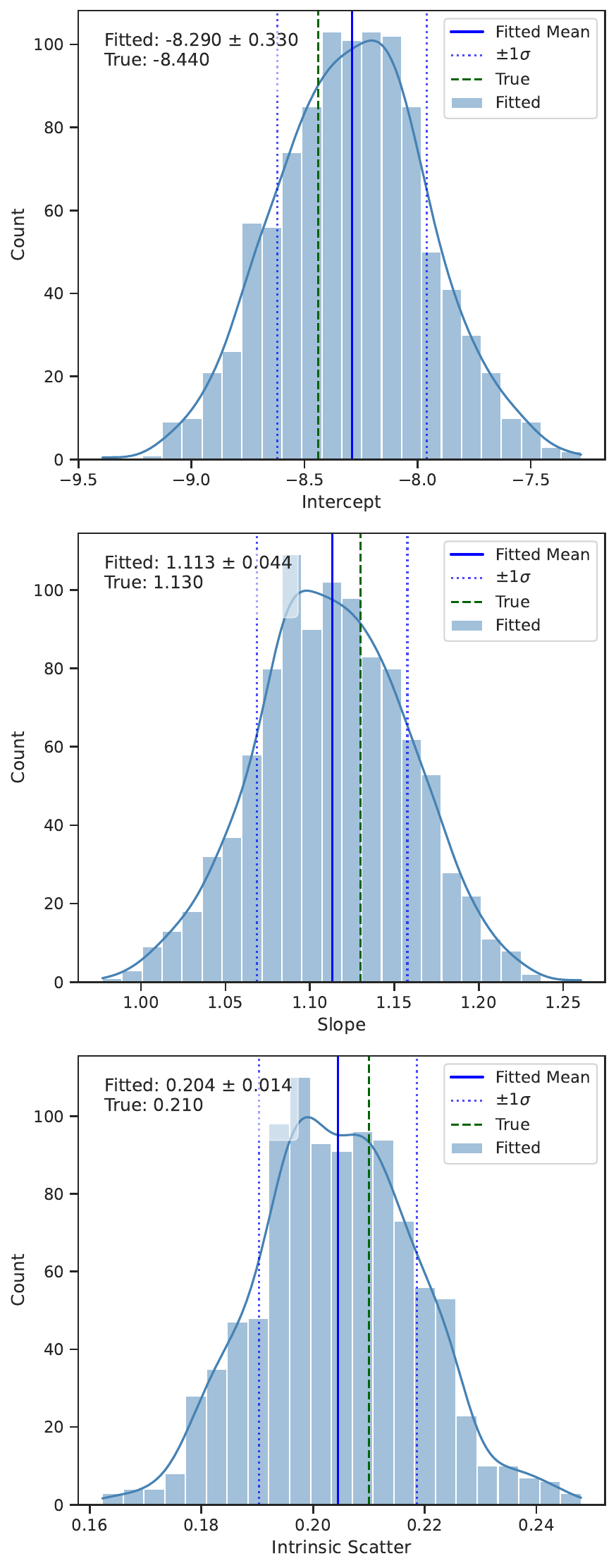}
\caption{Distributions of the fitted parameters obtained from 1000 simulated datasets (see Appendix\,\ref{app:NumSimBias}). From top to bottom : intercept, slope, and intrinsic scatter. The blue histograms show the results from our likelihood-based fitting method, with the solid and dotted blue lines indicating the mean and the $\pm 1\sigma$ intervals. The true input value used to generate the simulations is shown as a dashed green line. Each panel is annotated with the mean and standard deviation of the fitted parameter and the true input value in the top left corner.}
\label{fig:NumSimBias}
\end{figure}
\newpage

We then generate instrumental noise maps scaled to the noise properties of our ALMA data. We first convolve these noise maps to the same resolution, rescale their standard deviation, and added them to the gas maps. For the SFR maps, we applied the typical log-normal errors found by CIGALE for the real data. This process was repeated 13 times to produce a mock population comparable in size and complexity to our observed sample. Each mock dataset is subsequently analysed using our likelihood-based fitting procedure of Sect.\,\ref{sec:likelihood}, identical to what used for the real observations. We recover maximum likelihood solution of the slope, intercept, and intrinsic scatter parameters. The distributions of the fitted parameters across 1000 mock population realisations were then examined to assess the fidelity of the fitting method, see Fig.\,\ref{fig:NumSimBias}.

We find that, for each parameter, the true input value lies well within the $1\sigma$ confidence interval of the recovered distribution. Still, small systematic offsets between the mean fitted values and the true parameters are observed. The intercept is overestimated by approximately 0.15\,dex relative to the true value, while the slope and intrinsic scatter are underestimated by about 0.017 and 0.06\,dex, respectively. These relative differences correspond to roughly 0.45, 0.39, and 0.43 times the respective $1\sigma$ uncertainties, indicating that the deviations are minor compared to the statistical errors. We note that the $1\sigma$ intervals from the mock ensemble are systematically smaller than those from the real data. It could be due to a better S/N or the larger number of beams with a [CII] signal in our simulation. We therefore conclude that our approach does not systematically bias the estimates of the key parameters of the star formation relation.

\section{Alternative Fitting Methods}\label{app:otherfittingmethods}

In addition to this primary analysis, we use two other fitting methods to assess the validity of the likelihood model and its maximisation approach: the Bayesian linear regression package \texttt{linmix} and the MCMC sampler \texttt{emcee} \citep{foreman-mackey2013}.

For the MCMC analysis, we configure 32 walkers with 1,000 steps per chain, discarding the first 100 steps as burn-in and thinning chains by a factor of 20 to mitigate autocorrelation. To take into account the correlation between pixels, we normalise the likelihood from Eq.\,\ref{eq:Likelihood} by the number of independent pixels per beam. This approach aims to approximate the number of independent points per resolution element. However, since beam smoothing may effectively yield twice as many statistically independent noise samples as naively expected \citep[see][]{Condon1997,Condon1998}, we thus divide the likelihood by twice the number of pixels per beam. The impact of this normalisation is illustrated by the following MCMC fitting results. Without any likelihood normalisation, we obtain $\beta=0.79^{+0.01}_{-0.02}$ and $\sigma_i=0.17^{+0.00}_{-0.00}$. When normalising the likelihood by twice the number of pixels per beam, the results broaden to $\beta=0.78^{+0.09}_{-0.08}$ and $\sigma_i=0.17^{+0.02}_{-0.02}$. This demonstrates that accounting for the effective number of independent data points leads to larger and more realistic uncertainty estimates on the fitted parameters.

The \texttt{linmix} method uses Bayesian regression to account for measurement errors in both variables and intrinsic scatter, providing robust estimates of the slope, intercept, and intrinsic scatter of the studied [CII]-SFR relation. However, like in classical MCMC approach, it assumes uncorrelated pixels, ignoring spatial correlations within the beam and thus underestimate uncertainties in the presence of significant beam smoothing.

While both the \texttt{linmix} and MCMC fitting procedures provide estimates of the relation parameters and their uncertainties, they do not fully account for all sources of correlation, noise propagation, or population sampling effects present in our data. In particular, neither method can rigorously handle the spatial correlations and noise structure introduced by beam smoothing, nor the potential biases arising from the small sample size. For these reasons, we consider our primary resampling-based approaches described in Sect.\,\ref{sec:likelihood} and Sect.\,\ref{sec:resampling} to provide more reliable and conservative estimates of the uncertainties and best-fit parameters for the [CII]–SFR relation.

\onecolumn
\section{Other sources property maps}\label{app:othersourcesCIGALE}

\begin{figure}[h!]
    \centering
    \includegraphics[width=\textwidth]{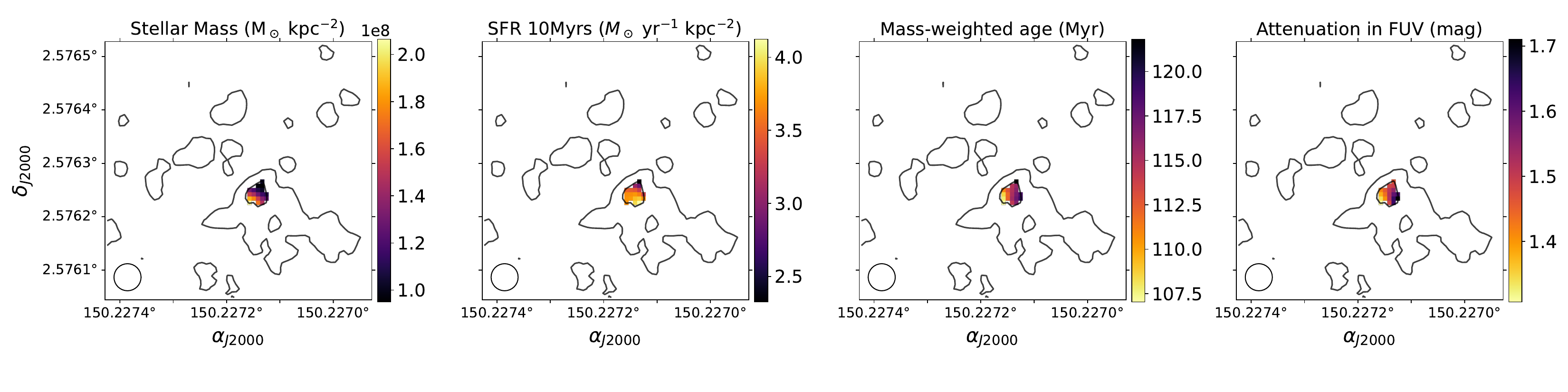}
    \includegraphics[width=\textwidth]{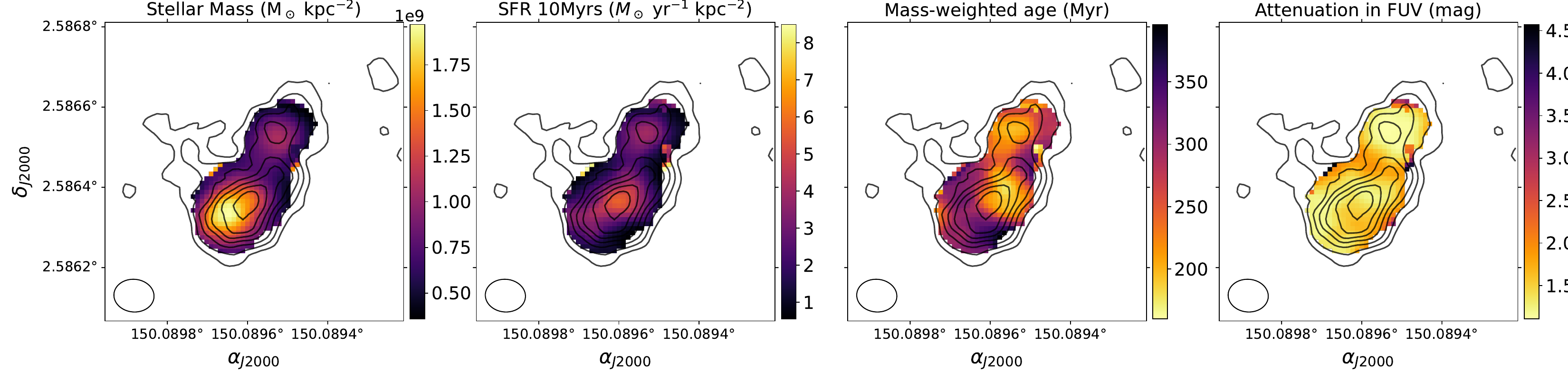}
    \includegraphics[width=\textwidth]{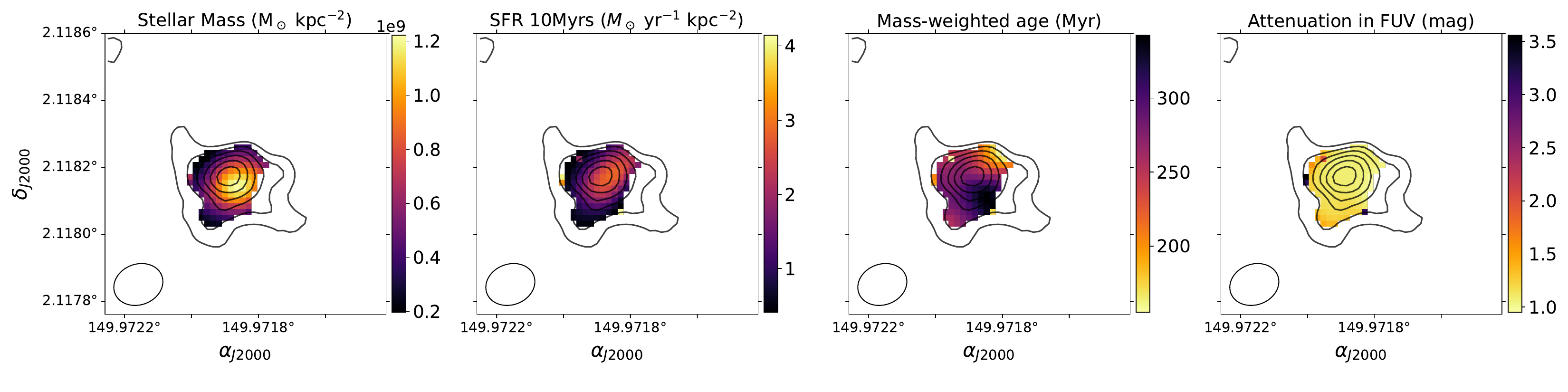}
    \includegraphics[width=\textwidth]{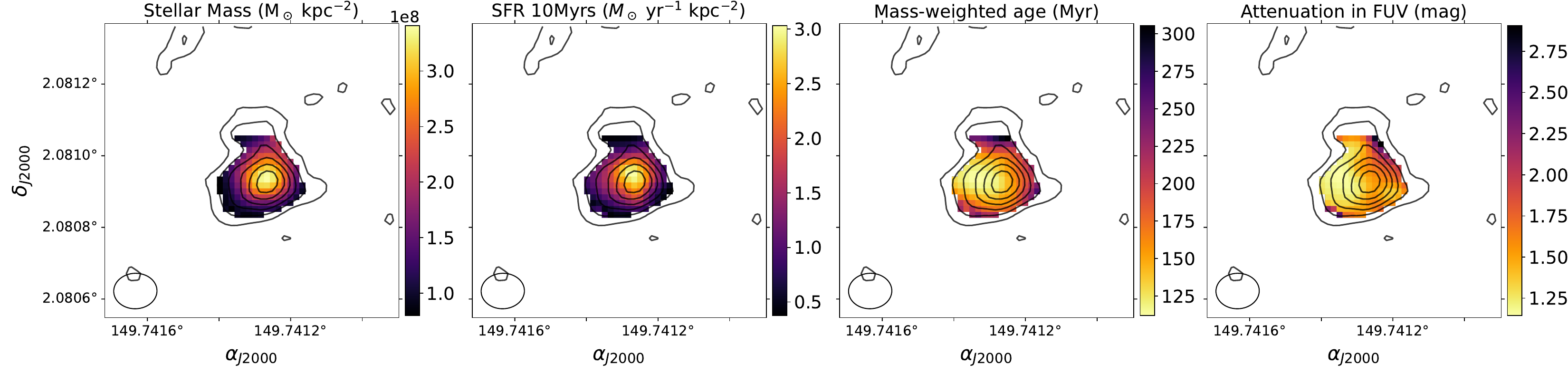}
    \caption{Similar to Fig.\,\ref{fig:CIGALEparams} for CRISTAL\_01, CRISTAL\_02, CRISTAL\_03, and CRISTAL\_04 (top to bottom)}
    \label{fig:C01-04CIGALE}    
\end{figure}

\begin{figure}
    \centering
    \includegraphics[width=\textwidth]{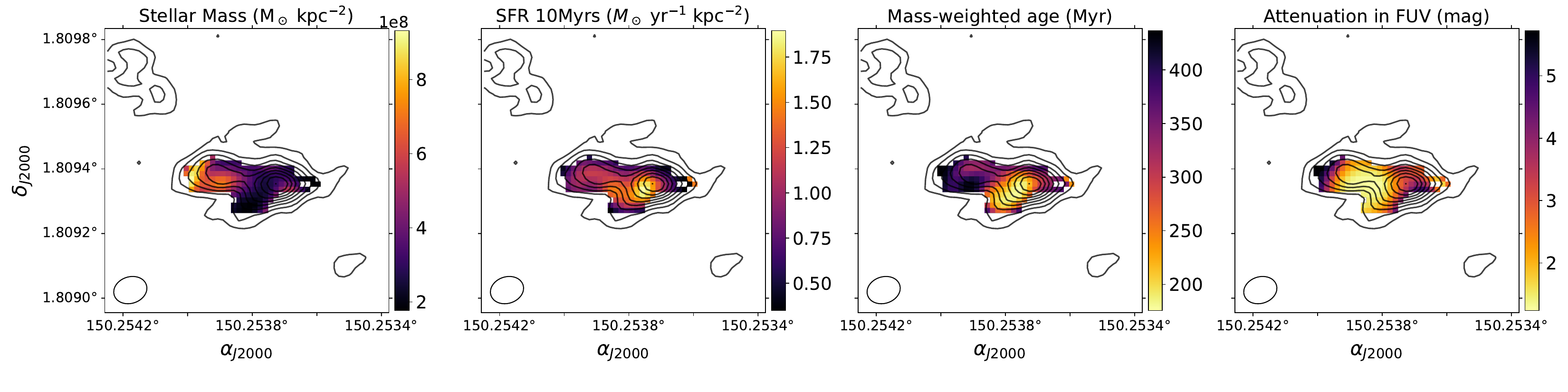}
    \includegraphics[width=\textwidth]{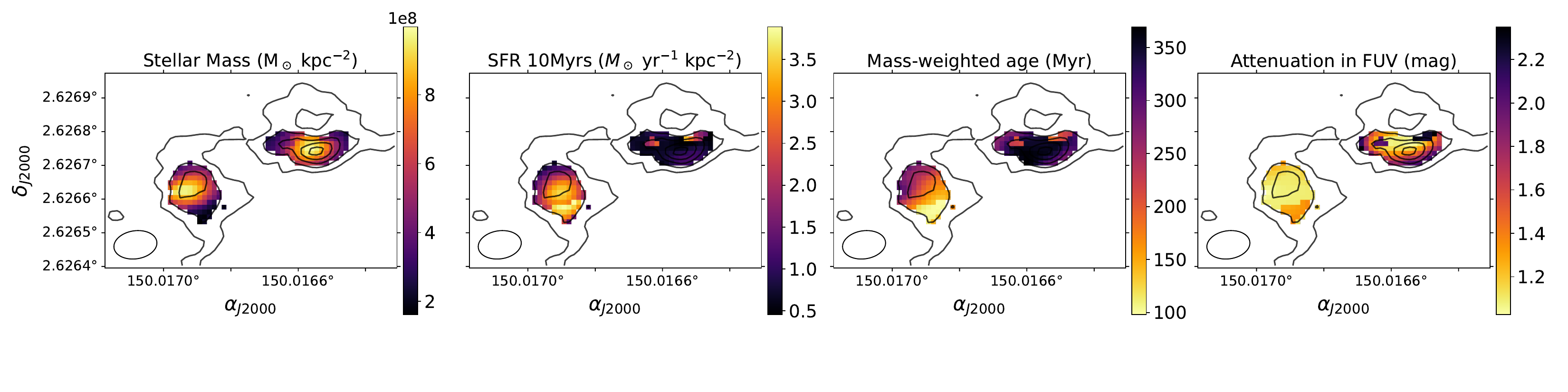}
    \includegraphics[width=\textwidth]{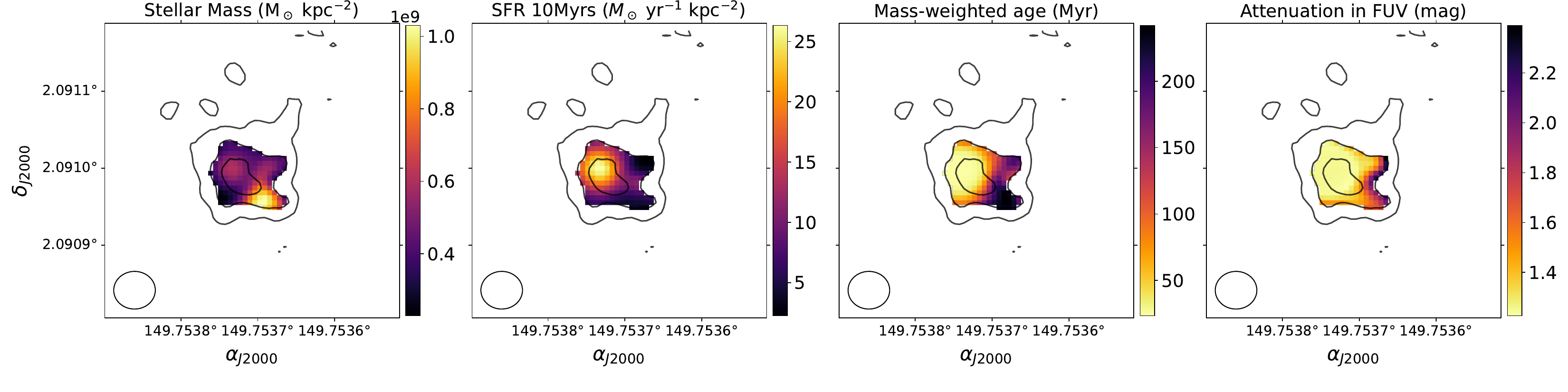}
    \includegraphics[width=\textwidth]{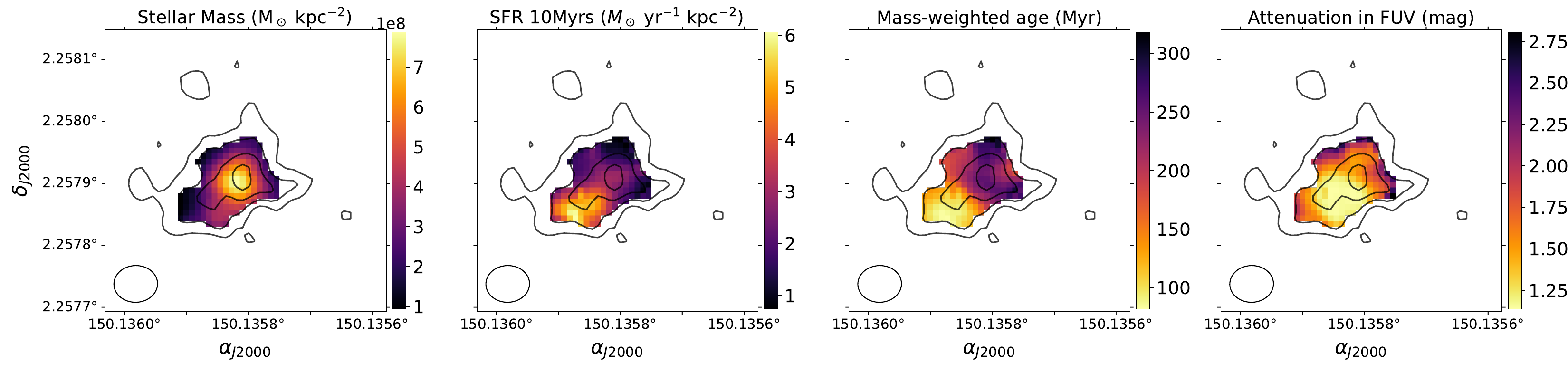}
    \includegraphics[width=\textwidth]{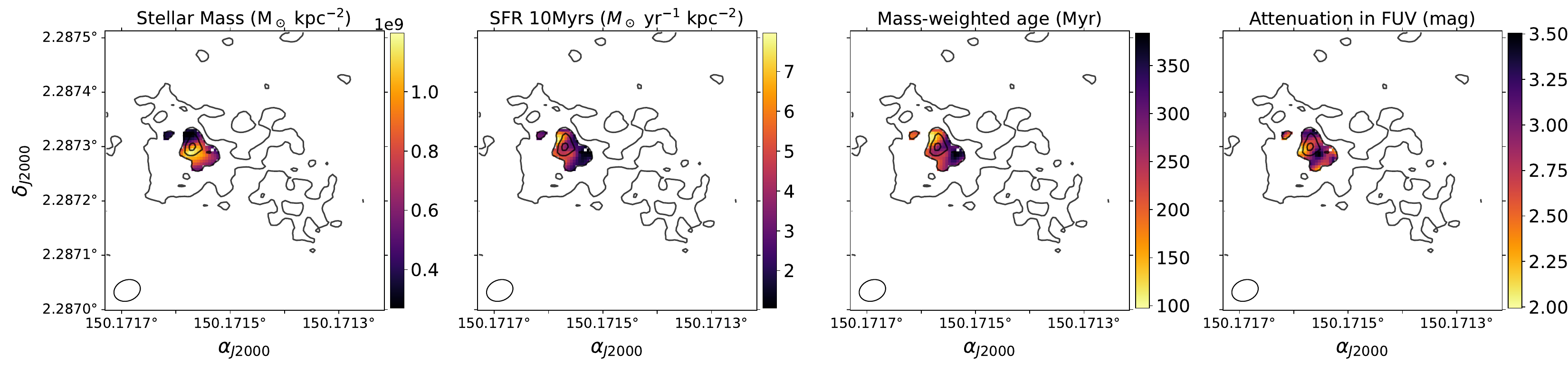}
    \caption{Similar to Fig.\,\ref{fig:CIGALEparams} for CRISTAL\_06, CRISTAL\_07, CRISTAL\_09, CRISTAL\_11 and CRISTAL\_13 (top to bottom)}
    \label{fig:C06-13CIGALE}    
\end{figure}

\begin{figure}
    \centering
    \includegraphics[width=\textwidth]{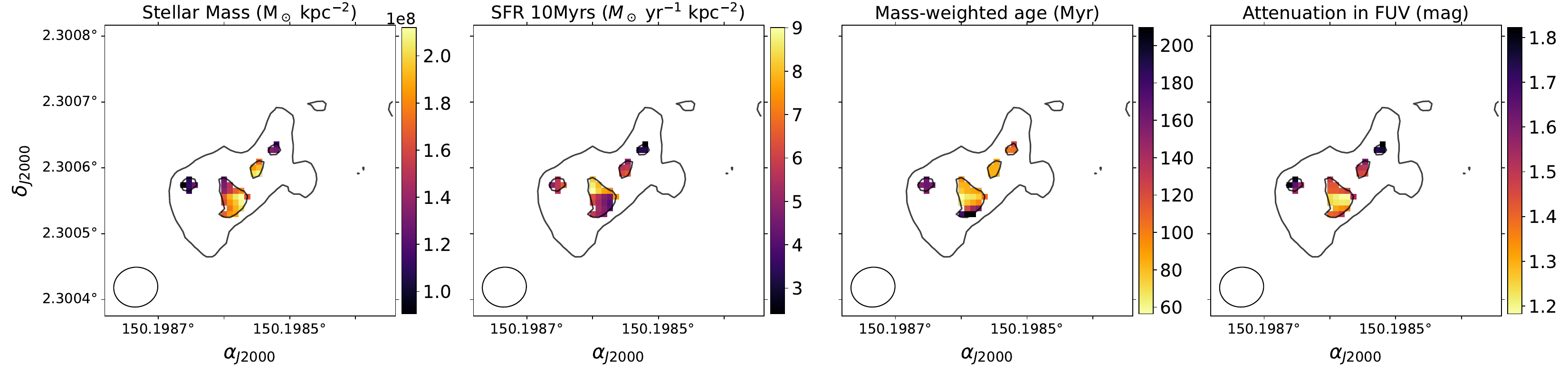}
    \includegraphics[width=\textwidth]{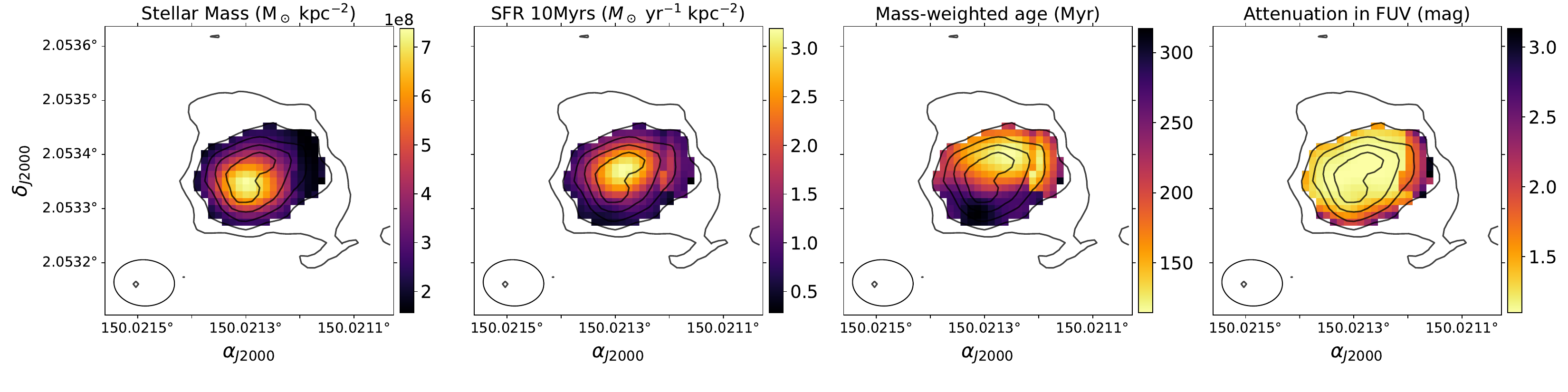}
    \includegraphics[width=\textwidth]{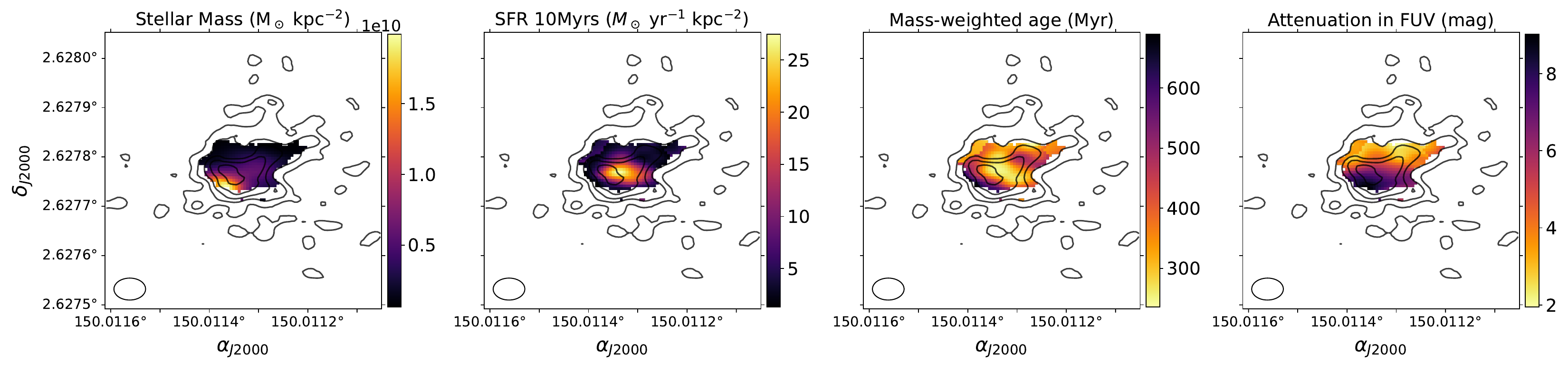}
    \caption{Similar to Fig.\,\ref{fig:CIGALEparams} for CRISTAL\_15, CRISTAL\_19, and CRISTAL\_24 (top to bottom)}
    \label{fig:C15-24CIGALE}    
\end{figure}

As in Fig.\,\ref{fig:CIGALEparams}, we show here example of spatially resolved physical parameters maps obtained with the CIGALE SED modelling for the other sources in the sample in Figs.\,\ref{fig:C01-04CIGALE},\ref{fig:C06-13CIGALE} and \ref{fig:C15-24CIGALE}. In each case, we have: stellar mass surface density ($\rm M_\sun\,kpc^{-2}$), star formation rate surface density averaged over 10\,Myrs ($\rm M_\sun\,yr^{-1}\,kpc^{-2}$), mass-weighted age (Myr) and attenuation in FUV rest-frame (mag). Results are shown with the final selection used for the [CII]-SFR and KS relation analysis, that is pixels with $\rm S/N\geq2$ in at least 3 photometric bands and where $\Sigma_{\rm [CII]}>5\sigma_{\rm[CII]}$. Across the sample, we find that the peaks of stellar mass and SFR are generally aligned. In contrast, regions of highest SFR tend to coincide with diminished FUV attenuation, and the mass-weighted age maps reveal that the oldest stellar populations are preferentially located in regions of highest stellar mass surface density.

\end{appendix}

\end{document}